\newcommand{\docstyle}{1} 
\ifnum\docstyle=0
\documentclass[final,5p,times,twocolumn]{elsarticle}
\else
\documentclass[10pt,journal,compsoc]{IEEEtran}
\fi

\usepackage{acronym}
\usepackage{graphicx}
\usepackage{tabularx}
\usepackage{enumerate}
\usepackage{color}
\usepackage{amsmath}
\usepackage{paralist}
\usepackage{threeparttable}
\usepackage{booktabs}
\usepackage{colortbl,hhline}
\usepackage{boldline}
\usepackage{colortbl}
\usepackage{array}
\usepackage{adjustbox}
\usepackage{multirow,tabu}
\usepackage{algorithm,algpseudocode}
\usepackage{tikz,pgfplots}
\usetikzlibrary{arrows}
\usepackage{pgfplotstable}
\usepgfplotslibrary{units,groupplots}
\usepackage{wrapfig}
\usepackage{varwidth,xcolor}
\usepackage{soul}
\usepackage{subcaption}
\usepackage{stfloats}
\usepackage{fancyhdr}

\newcommand{\highlightnewtext}{0}
\ifnum\highlightnewtext=1
\newcommand{\revised}[1]{{\leavevmode\color{blue}{#1}}}
\else
\newcommand{\revised}[1]{{\leavevmode\color{black}{#1}}}
\fi

\newcommand{\papertitle}{Introducing Packet-Level Analysis in Programmable Data Planes\\ to Advance Network Intrusion Detection}

\newcommand{\paperkeywords}{Programmable Data Planes, Network Intrusion Detection, Packet-Level features}

\newcommand{\mytinysize}{\fontsize{6}{7}\selectfont}
\pgfplotsset{
	compat = newest,
	tick label style={font=\sffamily\scriptsize},
	label style={font=\sffamily\scriptsize},
	legend style={font=\sffamily\mytinysize\raggedleft},
	legend cell align=left,
	grid style={dotted,gray}
}

\newcolumntype{?}{!{\vrule width 0.8pt}}

\definecolor{mygray}{RGB}{220,220,220}
\definecolor{skyblue}{RGB}{86,180,233}
\definecolor{bluish-green}{RGB}{0,158,115}
\definecolor{myblue}{RGB}{86,114,178}
\definecolor{vermilion}{RGB}{213,94,0}
\definecolor{reddish-purple}{RGB}{204,121,167}
\definecolor{deeplilac}{rgb}{0.6, 0.33, 0.73}
\definecolor{darkgreen}{rgb}{0.0, 0.2, 0.13}

\newlength{\Oldarrayrulewidth}

\graphicspath{{artworks/}{results/}{plots/}}
\DeclareGraphicsExtensions{.pdf,.jpeg,.png,.svg}
\ifnum\docstyle=0
	\journal{Computer Networks}
\fi

\begin{document}

\title{\papertitle}

\author{Roberto Doriguzzi-Corin$^\alpha$, Luis Augusto Dias Knob$^\alpha$, Luca Mendozzi$^\beta$, Domenico Siracusa$^\alpha$, Marco Savi$^\beta$\\
\small{$^\alpha$Cybersecurity Centre, Fondazione Bruno Kessler, Trento - Italy}\\
\small{$^\beta$University of Milano-Bicocca, Department of Informatics, Systems and Communication (DISCo), Milano - Italy}
}

\acrodef{acl}[ACL]{Access Control List}
\acrodef{ai}[AI]{Artificial Intelligence}
\acrodef{ahd}[AHD]{Abnormal Health Detection}
\acrodef{alu}[ALU]{Arithmetic Logic Unit}
\acrodef{ann}[ANN]{Artificial Neural Network}
\acrodef{api}[API]{Application Programming Interface}
\acrodef{bmv2}[BMv2]{Behavioral Model version 2}
\acrodef{bow}[BoW]{Bag-of-Words}
\acrodef{cnn}[CNN]{Convolutional Neural Network}
\acrodef{cpe}[CPE]{Customer Premise Equipment}
\acrodef{dl}[DL]{Deep Learning}
\acrodef{dlp}[DLP]{Data Loss/Leakage Prevention}
\acrodef{dpi}[DPI]{Deep Packet Inspection}
\acrodef{dnn}[DNN]{Deep Neural Network}
\acrodef{dns}[DNS]{Domain Name System}
\acrodef{dos}[DoS]{Denial of Service}
\acrodef{ddos}[DDoS]{Distributed Denial of Service}
\acrodef{dscp}[DSCP]{Differentiated Services Code Point}
\acrodef{ebpf}[eBPF]{extended Berkeley Packet Filter}
\acrodef{ewma}[EWMA]{Exponential Weighted Moving Average}
\acrodef{fl}[FL]{Federated Learning}
\acrodef{fpu}[FPU]{Floating Point Unit}
\acrodef{foss}[FOSS]{Free and Open-Source Software}
\acrodef{fnr}[FNR]{False Negative Rate}
\acrodef{fpr}[FPR]{False Positive Rate}
\acrodef{fvg}[\textsc{FedAvg}]{Federated Averaging}
\acrodef{gdpr}[GDPR]{General Data Protection Regulation}
\acrodef{gpu}[GPU]{Graphics Processing Unit}
\acrodef{ha}[HA]{Hardware Appliance}
\acrodef{ics}[ICS]{Industrial Control System}
\acrodef{ids}[IDS]{Intrusion Detection System}
\acrodef{iid}[i.i.d.]{independent and identically distributed}
\acrodef{ilp}[ILP]{Integer Linear Programming}
\acrodef{iot}[IoT]{Internet of Things}
\acrodef{isp}[ISP]{Internet Service Provider}
\acrodef{ips}[IPS]{Intrusion Prevention System}
\acrodef{jsd}[JSD]{Jensen-Shannon Distance}
\acrodef{ldap}[LDAP]{Lightweight Directory Access Protocol}
\acrodef{lstm}[LSTM]{Long Short-Term Memory}
\acrodef{lucid}[\textsc{Lucid}]{Lightweight, Usable CNN in DDoS Detection}
\acrodef{mau}[MAU]{Match-Action Unit}
\acrodef{mbgd}[MBGD]{Mini-Batch Gradient Descent}
\acrodef{mips}[MIPS]{Millions of Instructions Per Second}
\acrodef{ml}[ML]{Machine Learning}
\acrodef{mlp}[MLP]{Multi-Layer Perceptron}
\acrodef{mssql}[MSSQL]{Microsoft SQL}
\acrodef{nat}[NAT]{Network Address Translation}
\acrodef{netbios}[NetBIOS]{Network Basic Input/Output System}
\acrodef{nic}[NIC]{Network Interface Controller}
\acrodef{nids}[NIDS]{Network Intrusion Detection System}
\acrodef{nf}[NF]{Network Function}
\acrodef{nfv}[NFV]{Network Function Virtualization}
\acrodef{nsc}[NSC]{Network Service Chaining}
\acrodef{ntp}[NTP]{Network Time Protocol}
\acrodef{of}[OF]{OpenFlow}
\acrodef{ood}[o.o.d.]{out-of-distribution}
\acrodef{os}[OS]{Operating System}
\acrodef{ourtool}[P4DDLe]{P4-empowered Ddos detection with Deep Learning}
\acrodef{p4}[P4]{Programming Protocol-independent Packet Processors}
\acrodef{pess}[PESS]{Progressive Embedding of Security Services}
\acrodef{pisa}[PISA]{Protocol Independent Switch Architecture}
\acrodef{pop}[PoP]{Point of Presence}
\acrodef{portmap}[Portmap]{Port Mapper}
\acrodef{ppv}[PPV]{Positive Predictive Value}
\acrodef{ps}[PS]{Port Scanner}
\acrodef{qoe}[QoE]{Quality of Experience}
\acrodef{qos}[QoS]{Quality of Service}
\acrodef{rnn}[RNN]{Recurrent Neural Network}
\acrodef{rpc}[RPC]{Remote Procedure Call}
\acrodef{sdn}[SDN]{Software Defined Networking}
\acrodef{nn}[NN]{Neural Network}
\acrodef{bnn}[BNN]{Binary Neural Network}
\acrodef{sfnr}[sFNR]{system False Negative Rate}
\acrodef{sla}[SLA]{Service Level Agreement}
\acrodef{snf}[SNF]{Security Network Function}
\acrodef{snmp}[SNMP]{Simple Network Management Protocol}
\acrodef{ssdp}[SSDP]{Simple Service Discovery Protocol}
\acrodef{svm}[SVM]{Support Vector Machine}
\acrodef{tc}[TC]{Traffic Classifier}
\acrodef{tftp}[TFTP]{Trivial File Transfer Protocol}
\acrodef{tor}[ToR]{Top of Rack}
\acrodef{tos}[ToS]{Type of Service}
\acrodef{tpr}[TPR]{True Positive Rate}
\acrodef{tsp}[TSP]{Telecommunication Service Provider}
\acrodef{unb}[UNB]{University of New Brunswick}
\acrodef{vm}[VM]{Virtual Machine}
\acrodef{vne}[VNE]{Virtual Network Embedding}
\acrodef{vnep}[VNEP]{Virtual Network Embedding Problem}
\acrodef{vnf}[VNF]{Virtual Network Function}
\acrodef{vsnf}[VSNF]{Virtual Security Network Function}
\acrodef{vpn}[VPN]{Virtual Private Network}
\acrodef{xdp}[XDP]{eXpress Data Path}
\acrodef{wan}[WAN]{Wide Area Network}
\acrodef{waf}[WAF]{Web Application Firewall}

\ifnum\docstyle=0
\begin{abstract}
Programmable data planes offer precise control over the low-level processing steps applied to network packets, serving as a valuable tool for analysing malicious flows in the field of intrusion detection. Albeit with limitations on physical resources and capabilities, they allow for the efficient extraction of detailed traffic information, which can then be utilised by \ac{ml} algorithms responsible for identifying security threats. In addressing resource constraints, existing solutions in the literature rely on compressing network data through the collection of statistical traffic features in the data plane. While this compression saves memory resources in switches and minimises the burden on the control channel between the data and the control plane, it also results in a loss of information available to the \ac{nids}, limiting access to packet payload, categorical features, and the semantic understanding of network communications, such as the behaviour of packets within traffic flows.
This paper proposes \acsu{ourtool}, a framework that exploits the flexibility of P4-based programmable data planes for packet-level feature extraction and pre-processing.
\ac{ourtool} leverages the programmable data plane to extract raw packet features from the network traffic, categorical features included, and to organise them in a way that the semantics of traffic flows are preserved. To minimise memory and control channel overheads, \ac{ourtool} selectively processes and filters packet-level data, so that \revised{only the} features required by the \ac{nids} are collected. 
The experimental evaluation with recent \ac{ddos} attack data demonstrates that the proposed approach is very efficient in collecting compact and high-quality representations of network flows, ensuring precise detection of \ac{ddos} attacks.

\end{abstract}

\maketitle
\else
\maketitle
\begin{abstract}
Programmable data planes offer precise control over the low-level processing steps applied to network packets, serving as a valuable tool for analysing malicious flows in the field of intrusion detection. Albeit with limitations on physical resources and capabilities, they allow for the efficient extraction of detailed traffic information, which can then be utilised by \ac{ml} algorithms responsible for identifying security threats. In addressing resource constraints, existing solutions in the literature rely on compressing network data through the collection of statistical traffic features in the data plane. While this compression saves memory resources in switches and minimises the burden on the control channel between the data and the control plane, it also results in a loss of information available to the \ac{nids}, limiting access to packet payload, categorical features, and the semantic understanding of network communications, such as the behaviour of packets within traffic flows.
This paper proposes \acsu{ourtool}, a framework that exploits the flexibility of P4-based programmable data planes for packet-level feature extraction and pre-processing.
\ac{ourtool} leverages the programmable data plane to extract raw packet features from the network traffic, categorical features included, and to organise them in a way that the semantics of traffic flows are preserved. To minimise memory and control channel overheads, \ac{ourtool} selectively processes and filters packet-level data, so that \revised{only the} features required by the \ac{nids} are collected. 
The experimental evaluation with recent \ac{ddos} attack data demonstrates that the proposed approach is very efficient in collecting compact and high-quality representations of network flows, ensuring precise detection of \ac{ddos} attacks.

\end{abstract}

\fi

\thispagestyle{fancy}
\renewcommand{\headrulewidth}{0pt}
\chead{\scriptsize This is the author's version of an article that has been published in Elsevier Computer Networks. Changes were made to\\this version by the publisher prior to publication. The final version of record is available at {\color{blue}{https://doi.org/10.1016/j.comnet.2023.110162}}. \\ The source code associated with this project is available at {\color{blue}{https://github.com/risingfbk/p4ddle}}.}
\cfoot{\scriptsize Copyright (c) 2024 Elsevier. Personal use is permitted. For any other purposes, permission must be obtained from Elsevier by contacting the Permissions Helpdesk Support Center {\color{blue}{https://service.elsevier.com/app/contact/supporthub/permissions-helpdesk/}}.}

\section{Introduction}\label{sec:introduction}
Network intrusions and anomalies are one of the most significant plagues in modern communication networks. 
As the number and complexity of incidents are constantly increasing \cite{enisa-report}, it has become imperative to implement meticulous monitoring and robust counteraction measures to effectively detect and mitigate these threats.
In the past decade, network monitoring has lived a second youth, thanks in large part to the prominence of \ac{sdn} \cite{rojas2018we, khorsandroo2021hybrid} and to the rise of \ac{ml} technologies in networking \cite{kanakis2022machine}. 
Within the network security domain, the fusion between the centralised control plane of \ac{sdn} and data-driven threat detection powered by \ac{ml} algorithms has proved remarkable efficacy in promptly identifying and mitigating network intrusions and anomalies \cite{sultana2019survey}. 

Despite the undeniable benefits brought by the synergy between \ac{sdn} and \ac{ml}, 
the implementation of robust network security solutions remains a challenge, primarily due to the fine-grained traffic features required by \ac{ml} algorithms. 
In this context, a direct and effective approach for leveraging the centralised control plane of \ac{sdn}, while ensuring that \ac{ml} algorithms receive the necessary traffic information, is through a technique known as \textit{packet mirroring} \cite{10.5555/3323234.3323252}. By employing packet mirroring, a duplicate copy of the network traffic is transmitted from the data plane to the control plane, where it undergoes traffic feature extraction and pre-processing before \ac{ml} algorithms can be executed upon it (Figure \ref{fig:packet_mirroring}).
Unfortunately, the channel between data and control planes often comes with severe bandwidth and latency bottlenecks \cite{9259415,9960816,9796830,9829799}. These inherent limitations imply that it may be overwhelmed by the sheer volume of mirrored data \cite{buhler2023enhancing}, thereby jeopardising ordinary \ac{sdn} network operations and compromising prompt response to ongoing network attacks.

The recent emergence of programmable data planes \cite{10.1145/3447868} has introduced a technology that offers a promising solution to tackle the above challenges. Programmable data planes enable the customization of the data plane pipeline (known as \textit{fastpath} \cite{xing2020netwarden}) using domain-specific languages like P4 \cite{10.1145/2656877.2656890}. This level of programmability and flexibility empowers network practitioners to optimise feature extraction, data pre-processing, and \ac{ml} inference operations, which can be offloaded to the data plane (see Figure \ref{fig:all_data_plane}):
with programmable data planes, it thus becomes possible to finely manipulate the type and volume of traffic data forwarded to the control plane (known as \textit{slowpath} \cite{xing2020netwarden}). 
However, it is important to note that the data plane has inherent limitations, such as a limited set of available arithmetic instructions and memory capacity in the match-action tables \cite{9110257}. As a result, only simple \ac{ml} models can be effectively offloaded, which may lead to noticeable degradation in inference performance \cite{xiong19}. 

We advocate leveraging the full potential of the centralised \ac{sdn} control plane by executing \ac{ml} inference directly on the \ac{sdn} controller. This approach allows network monitoring operations to exploit the advantages of a comprehensive, global view of the network, while also enabling the implementation and execution of sophisticated \ac{ml} algorithms. By performing \ac{ml} inference on the controller, the network can benefit from enhanced analysis and decision-making capabilities, leveraging the rich information available at the control plane.
Simultaneously, we recognise the value of programmable data planes in performing traffic feature extraction and pre-processing. By leveraging the programmability of data planes, it becomes possible to precisely control the amount of data that traverses the control channel. This approach allows for selective processing and filtering of data at a data plane level, reducing the burden on the control channel and enabling efficient resource allocation (see Figure \ref{fig:preprocessing_control_plane}).

This concept has been extensively explored in the scientific literature, with several studies proposing solutions based on the aggregation of traffic statistics computed within the data plane. The primary objective of these solutions is to optimise the load on the control channel and effectively manage memory utilization in the data plane \cite{musumeci2022machine,macias2021oracle,flowlens,zang2022sdn,roshani2022hybriddad}.
One notable drawback of relying solely on statistical traffic features, such as flow duration, averages, maximums, minimums, and standard deviations of attributes like packet size, rate, and inter-arrival time, is the inability to report categorical features (e.g., IP and TCP flags, \revised{\ac{dscp}, etc.), the timestamp of packets} or portion of packets' payload directly to the control plane.  
These packet-level traffic features are crucial \revised{to detect some types of network intrusions, such as brute force attacks (e.g., by monitoring the TCP flags, including SYN, ACK, FIN and RST flags \cite{hofstede2014ssh}), anomalous \acs{qos} settings (monitoring the \ac{dscp} code points) \cite{custura2018exploring}, TCP, UDP and ICMP fragmentation attacks (monitoring the IP flags) \cite{illy2022ml}, etc. In addition, the timestamp of packets and portions of the payload are employed in combination with categorical features by} certain \ac{ml}-based \acp{nids} in the current state of the art (e.g., \cite{lucid,xu2020method,alani2022botstop,tr-ids,han2023network}, among many others) to learn the semantic of malicious traffic flows and to segregate them from legitimate network activities.

\begingroup
\setlength{\tabcolsep}{1pt} 
\begin{figure*}[t!]
    \begin{tabular}{ccc}
		\begin{subfigure}{0.33\linewidth}
            \begin{center}
                \includegraphics[width=1\linewidth]{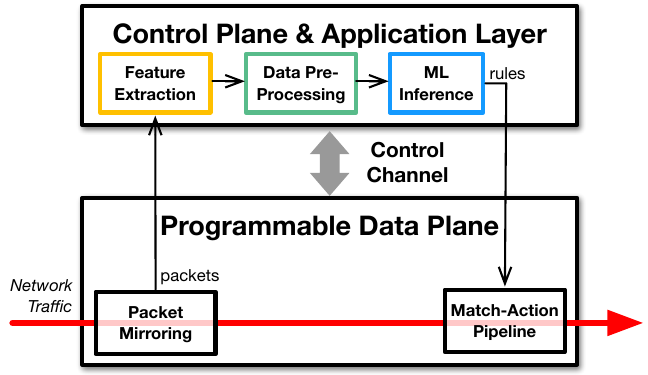}
                \caption{Packet mirroring.}
				\label{fig:packet_mirroring}
            \end{center}
        \end{subfigure} 
		&
        \begin{subfigure}{0.33\linewidth}
            \begin{center}
                \includegraphics[width=1\linewidth]{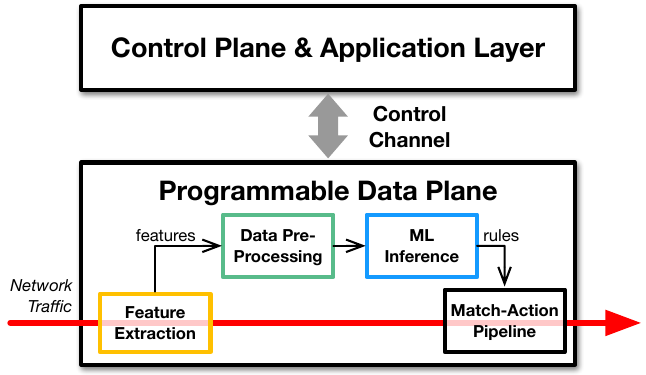}
                \caption{In-network traffic monitoring.}
				\label{fig:all_data_plane}
            \end{center}
        \end{subfigure} 
		& 
		\begin{subfigure}{0.33\linewidth}
            \begin{center}
                \includegraphics[width=1\linewidth]{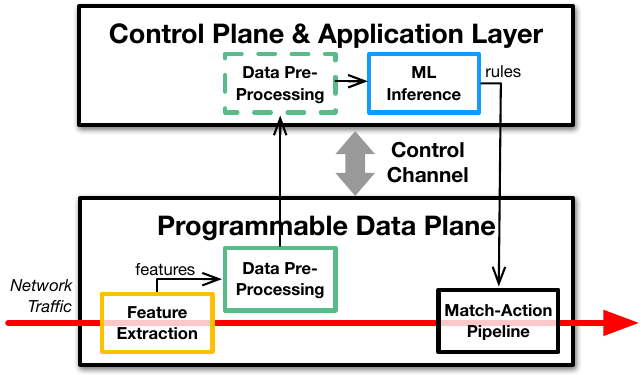}
                \caption{Hybrid approach.}
				\label{fig:preprocessing_control_plane}
            \end{center}
        \end{subfigure} 
		\\
    \end{tabular}
	\caption{Approaches to traffic monitoring with programmable data planes.}
	\label{fig:processing_pipeline}
\end{figure*}
\endgroup

In this paper, we propose \acs{ourtool} (\aclu{ourtool}), a framework for efficient packet-level feature extraction and pre-processing in P4-based programmable data planes. 
\ac{ourtool} inherently supports the categorical features, and it has been designed to preserve the semantic integrity of the flows.
With these properties, \ac{ourtool} ensures that the \ac{ml}-based \ac{nids} executed in the control plane can capture the relevant protocols, data formats, application behaviours and traffic patterns, allowing for a comprehensive understanding of the network traffic semantics.

\ac{ourtool} exploits stateful memory objects (\textit{P4 registers}) and a \textit{counting Bloom filter} to efficiently select and store the packet-level features that are relevant for the \ac{ml} model.
This approach enables the data plane to discard redundant data,  which would otherwise be disregarded by the control plane, with great benefits in terms of control channel usage and control plane processing load. The selected features are stored within two ring buffers, which have been specifically designed to avoid race conditions between the control plane (reading the stored features) and the data plane (writing them). The latter aspect has been often neglected by previous works.

To the best of our knowledge, this is the first work addressing the challenges of packet-level feature extraction in the data plane for network security applications.
In contrast to existing works, \ac{ourtool} combines the flexibility of packet-level features with the network-wide view provided to the \ac{nids} by the centralised \ac{sdn} control plane (not available when the \ac{nids} runs completely in the data plane \cite{coelho2022,qin2020}). We believe that the combination of these two features sets \ac{ourtool} apart from current state-of-the-art approaches to network attack detection with programmable data planes.

We have extensively evaluated \ac{ourtool} using a recent dataset of \ac{ddos} attacks and a well-known \ac{nids} based on a \ac{cnn} that takes packet-level features as input \cite{lucid}. 
We have compared our approach against a naïve strategy for packet-level feature extraction, which stores the data in the buffers sequentially without any optimisation logic.
In a comprehensive range of testing scenarios, we empirically show that \ac{ourtool} is able to collect more feature-rich traffic flows. As a result, our approach achieves a significantly lower system \ac{fnr}, measured as the sum of overlooked and misclassified malicious flows, enabling faster detection and mitigation of network attacks.

\revised{
The main contributions of this work can be summarised as follows:
\begin{enumerate}
    \item A framework for P4-based programmable data planes called \ac{ourtool}, designed to cope with the requirements of modern \acp{nids}, which often rely on raw packet-level features for accurate cyber threat detection.
    \item A mechanism based on counting Bloom filters to efficiently group packet-level features into traffic flows while minimizing the chances of collisions between the flow identifiers.
    \item A comprehensive assessment of the proposed approach, including a sensitivity analysis of the Bloom filter size (directly linked to the probability of collisions), and the evaluation of a state-of-the-art \ac{nids} in detecting realistic \ac{ddos} attacks using \ac{ourtool}'s packet-level representation of the network traffic.
\end{enumerate}
}

The remainder of the paper is structured as follows. Section \ref{sec:related} reviews and discusses the related work. Section \ref{sec:background} provides an overview of P4 data plane programming, counting Bloom filters and the \ac{cnn} used in our experiments. Section \ref{sec:architecture} presents \ac{ourtool}'s architecture and the methods for feature extraction and storage. Section \ref{sec:setup} details the setup of the experimental evaluation presented in  Section \ref{sec:evaluation}.  Finally, the conclusions are given in \revised{Section} \ref{sec:conclusions}.

\section{Related Work}\label{sec:related}
One of the primary challenges in network traffic monitoring is finding a balance between the accuracy of the monitored traffic attributes and the level of resources (both network and computing) required to achieve it. 
In this regard, programmable network devices can be exploited to handle the monitoring operations in the data plane, effectively reducing the burden on both control plane and control channel. Nevertheless, despite the adaptability of modern data plane implementations, including those relying on the \ac{p4} programming language, they still have certain limitations in the types of operations they can perform. 
To address this issue, a common technique is to offload some of the computation from the data plane to the control plane (e.g., performing \ac{ml} inference), where more computing power and advanced tools are accessible. Nevertheless, this approach could be constrained by the limited capacity of the communication/control channel linking the data and control planes.

This section provides an overview of recent research that has tackled these challenges, with a particular emphasis on the network security domain, where achieving a balance between resource utilization and monitoring accuracy is critical for quickly identifying and responding to cyber threats. 

\subsection{ML-based In-network Traffic Classification}
In network monitoring applications, an effective solution to prevent the control channel from being overloaded with network data is to offload the tasks of traffic feature extraction, data pre-processing, and \ac{ml} inference to the programmable data plane (see Figure \ref{fig:all_data_plane}). This approach, also known as the \textit{in-network} approach, offers several advantages in terms of control channel utilization by minimising the interaction between data and control planes. However, despite its benefits, this approach also presents some significant drawbacks that should be taken into account. 

It is worth noting that the \ac{p4} programming language does not support floating-point operations and divisions \cite{ding2021tracking}. 
Consequently, some highly effective \ac{ml} algorithms, including \acp{ann}, cannot be directly implemented in the data plane, as those algorithms rely on model weights that are usually represented as floating-point numbers. 
Recent initiatives have been trying to enable floating-point operations to \ac{p4}, either with the adoption of dedicated hardware \acp{fpu} \cite{graham2020scalable} or by proposing hardware changes to the data plane's architecture of programmable devices \cite{yuan2022unlocking}. Although promising, these approaches present limited portability to existing \ac{p4} devices (switches and/or smartNICs), and may require a non-negligible time horizon for their adoption in more recent products.

For this reason, most of the existing proposals for \ac{ml}-based in-network traffic classification that exploit programmable data planes adopt simple \ac{ml} models, such as decision trees \cite{xiong19}\cite{zhang21}\cite{xavier2022map4}\cite{zheng2022}\cite{zhou2023efficient}, binary decision trees \cite{9796936}, random forests \cite{busse2019}\cite{lee2020}\cite{zheng2022}\cite{coelho2022}\cite{zhou2023efficient}, \ac{svm} \cite{xiong19}\cite{zheng2022}, Naïve Bayes \cite{xiong19}\cite{zheng2022}, K-means \cite{xiong19}\cite{zheng2022}, XGBoost \cite{zheng2022}. Concerning \ac{nn} models, \ac{bnn} can be successfully offloaded \cite{qin2020}\cite{siracusano22} thanks to their simplicity, as model weights are binary values. 

In all these works, the \ac{ml} models are implemented by exploiting either match-action tables, populated by the control plane with appropriate rules, or by making use of P4 registers, i.e., by hard-coding the model into the P4 program. The former case is more flexible, as the model can be updated by the control plane at runtime (e.g. after re-training) by just injecting new rules, but it consumes memory that is typically dedicated to traffic forwarding. 
In a recent work, Razavi et al. \cite{razavi2022} have implemented an \acp{ann} directly in the data plane. 
A notable limitation of this approach is the encoding of floating-point weights as 8-bit integers. As no performance evaluation is presented, the efficacy of the proposed solution remains uncertain.
In addition, some works \cite{zheng2022automating}\cite{swamy2022homunculus} focus on the design and implementation of frameworks for offloading to the data plane the most appropriate \ac{ml} model according to the task to be executed.

In general, what we can conclude by analysing the aforementioned papers is that the performance of \ac{ml}-based in-network traffic classification is often hindered by the inherent limitations of the \ac{p4} language and/or of the underlying hardware. This can make it challenging to offload \ac{ml} models with satisfactory classification performance.

In contrast to the solutions mentioned above, we have opted to execute the traffic classifier in the control plane. This approach allows us to choose the most appropriate \ac{ml} model for a given task, without being constrained by the limitations of the P4 language or the hardware of the devices. 
By leveraging the flexibility of the control plane, we can use more advanced \ac{ml} models to achieve better accuracy and performance as needed.
Furthermore, the centralised nature of the \ac{sdn} control plane enables the \ac{nids} to leverage network-wide traffic data, facilitating more accurate and comprehensive detection of network threats. This advantage is not attainable with the distributed inference employed by \textit{in-network} approaches.

\subsection{Interaction between Control Plane and Data Plane}
Offloading a portion of traffic processing to the control plane (Figure \ref{fig:preprocessing_control_plane}) requires careful consideration of the limitations imposed by the control channel in terms of bandwidth and latency. This consideration becomes even more crucial when a substantial amount of data must be exchanged between the two planes, as is for instance required when executing \ac{ml} inference in the control plane for \ac{ddos} attack detection \cite{lucid}.

NetWarden \cite{xing2020netwarden} is a defense solution designed to mitigate network covert channels, utilising programmable data planes. In order to achieve this goal, NetWarden's approach involves restricting the data plane to performing per-packet operations exclusively on header fields. On the other hand, the control plane is only used for batch operations, reducing the need for interaction between the control and data planes and optimising the overall performance of the system. DySO \cite{9829799} is a monitoring framework that shares the same fundamental principles as NetWarden, seeking to eliminate the bottleneck that can occur between data and control planes. Our research similarly aims to minimise the interaction between the two planes, accomplishing this by conducting feature extraction and selection in the data plane. By only forwarding essential data to the control plane for input to the \ac{ml} model, we can significantly reduce the overall system latency and optimise performance.

A couple of recent works take different approaches to solve the bottleneck issue with the control channel. Escala \cite{9796830} operates under the assumption that multiple control channels may be in use and may become overloaded. To address this issue, Escala offers the ability to elastically scale these channels at runtime as necessary, and to seamlessly migrate event streams (i.e., data transmitted from the data to the control plane) between different channels.
Chen et al. \cite{9960816} introduce MTP, a novel framework designed to optimise the placement of measurement points in the data plane, with the aim of minimising the overall cost associated with monitoring servers and network devices. This framework is based on an \ac{ilp} formulation, which defines a constraint on the capacity of the links between the data and control planes. In general, these works are orthogonal to our proposal and could be adopted to further alleviate the performance bottleneck of the control channel.

Finally, Bermudez Serna et al. \cite{9964491} focus on security aspects. They propose a reactive configuration mechanism that can be adopted to counteract attacks aimed at overloading the control channel which, given its constrained capacity, could easily get saturated by malicious traffic. Also, this strategy is orthogonal to our proposal and could be adopted together with it for enhanced security.

\subsection{Efficient Feature Extraction in the Data Plane}
Efficient feature extraction in programmable data planes is a challenging problem that has received limited attention in the scientific literature. The goal is to generate features that are compatible with the input layer of a \ac{ml} classifier, which is executed in the control plane for traffic monitoring or classification. Despite the importance of this problem, there are only a few studies that have addressed it and most of them, like our work, focus on \ac{ddos} detection as a use case.

In this respect, FlowLens \cite{flowlens} is a flow classifier designed for programmable switches that implements a mechanism for optimising the amount of data to be stored in memory and transferred through the control channel. 
While this feature enables efficient processing of network flows, race conditions on the shared memory are managed by the control plane, which deletes the flow tables to avoid concurrent read/write operations on the registers that store the traffic features. According to the paper, the flow tables are left empty until the reading process is completed. As a result, any incoming packets during this period cannot be processed or collected.

Musumeci et al. \cite{musumeci2022machine} propose an approach to SYN Flood \ac{ddos} attack detection that leverages \ac{p4}-based programmable data planes for feature extraction and pre-processing. The \ac{p4} program extracts statistical traffic features computed over a pre-defined time window, such as average packet length, percentage of TCP and UDP packets and percentage of TCP packets with active SYN flag. 
A similar approach is adopted by Zang et al. \cite{zang2022sdn}, who propose a \ac{ddos} detection system based on ensemble learning and data-plane feature extraction and pre-processing. 
Due to the reliance on global statistical features rather than per-flow features, both approaches are limited to producing binary outputs indicating the presence of an ongoing attack. However, their \ac{ml} classifiers cannot identify specific malicious flows.
HybridDAD \cite{roshani2022hybriddad} is another solution based on statistical features. In this case, the output of the \ac{ml} algorithm is a label that indicates whether there has been an attack during the previous time window, plus the class of the attack (among four types of DDoS flood attacks).

ORACLE \cite{macias2021oracle} implements a flow-based feature extraction and pre-processing in the data plane for the detection of \ac{ddos} attacks, which is offloaded to \ac{ml} algorithms executed in the control plane. 
To accomplish this, ORACLE employs a \ac{p4} program that collects per-flow statistical features, including flow duration, standard deviation of inter-arrival time, average packet size, and standard deviation of packet length. 
Apart from the inherent complexity of calculating statistical features in the data plane due to the limitations of the P4 language, ORACLE utilises hashing to index flows. With this approach, packets from different flows can be grouped together in the case of collisions. As a result, the computed statistics may be incorrect, making the final flow statistics unreliable and unusable.

\vspace{2mm}
Our solution addresses the limitations of prior research by implementing a double-block storage mechanism in the data plane, which effectively prevents race conditions. We achieve this by using two separate blocks (implemented using a set of \ac{p4} registers) and switching between them using a dedicated \ac{p4} register. The data plane writes to one block while the control plane reads from the other, ensuring consistency and minimising conflicts. Compared to previous methods based on statistical flow-level features, our approach leverages a packet-level representation of network traffic, which preserves the semantics of flows (i.e., the behaviour of packets within each flow) and the categorical features of packets (e.g., TCP flags, ICMP type, etc.). This information is crucial for various \ac{ml}-based \acp{nids} \cite{lucid,xu2020method,alani2022botstop,tr-ids,han2023network}.
 On the other hand, packet-level features may entail a greater amount of data transmission via the control channel, compared to flow-level statistical features. We alleviate this by proposing a \textit{flow-based} storage mechanism, that relies on a statistical technique called \textit{counting Bloom filter}~\cite{tarkoma2011theory}. By doing so, we can keep in memory only the essential data required for control plane operations, reducing the need to transmit extra traffic information that would ultimately be discarded.

\section{Background}\label{sec:background}

\subsection{Data Plane Programming with P4}\label{sec:p4_programming}

\ac{p4} is a language for expressing how the network traffic is processed by the data plane of programmable network elements such as hardware and software switches. \ac{p4} is target-independent, thus it supports a variety of targets such as ASICs, FPGAs, NICs and software switches. 

\begin{figure}[h!]
	\begin{center}
		\includegraphics[width=1\linewidth]{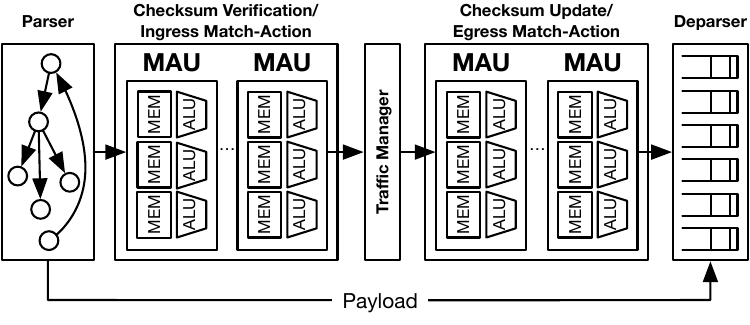}
		\caption{\acs{p4}-programmable pipeline components \revised{(\textit{V1model})}.}
		\label{fig:p4_design}
	\end{center}
\end{figure}

The \revised{basic} components of a \ac{p4}-programmable \revised{forwarding} pipeline are illustrated in Figure  \ref{fig:p4_design}.
\revised{This diagram outlines the \textit{Version 1.0 Switch Architecture Model} \cite{manzanares2021passive}, abbreviated as \textit{V1model}, which is widely adopted by developers for \ac{p4} program implementation. We utilised this model as a reference for \ac{ourtool}.
The pipeline encompasses} a \textit{Parser}, a state machine that extracts packet headers and metadata from the incoming bitstream, a \textit{Checksum verification} control block, an \textit{Ingress Match-Action} processing control block, an \textit{Egress Match-Action} processing control block, a \textit{Checksum update} control block and a \textit{Deparser}, which assembles the headers with the payload received from the Parser.

As shown in the figure, the Match-Action block may consist of one or multiple \acp{mau}. Each \ac{mau} comprises TCAM and SRAM memory blocks (MEM in the figure) for various purposes such as match-action tables, registers, meters, etc. The \acp{alu} blocks instead execute arithmetic operations and modifications on the packets' headers and metadata based on the content of the match-action tables\revised{, which host the traffic forwarding rules}. \acp{alu} may use stateful objects stored in the SRAM memory (e.g., registers) for additional operations.

Relevant to this work are the stateful objects of type \textit{register}. Registers belong to a wider category of stateful elements called \textit{extern objects}, which also include counters and meters. Unlike match-action tables\revised{, which are stored in the TCAM memory and} can be modified only by the control plane, extern objects \revised{in the SRAM memory} can be read and written from both data and control planes. Therefore, the access to registers, and to extern objects in general, should be managed with atomic operations to prevent race conditions.

In this work, we use registers to store per-packet and per-flow data that is written by the data plane and periodically read by the control plane. We initially enclosed read/write operations into atomic blocks to avoid race conditions between data and control planes. However, as we noticed that this approach causes a considerable loss of traffic data when the control plane locks the registers for reading, we adopted a more efficient strategy consisting of using two different registers, one for writing and one for reading. A third register serves as a switch to orchestrate the access to those two registers. Section \ref{sec:architecture} provides more details on this mechanism.

\subsection{Counting Bloom filters}\label{sec:counting-bloom-filter}
A \textit{counting Bloom filter} \cite{tarkoma2011theory} is an array of $m$ cells that utilizes probabilistic hashing techniques to keep an approximate count of items. We use a counting Bloom filter to efficiently count the number of packets belonging to the same flow that have been collected in a specific time frame.
The key idea behind a counting Bloom filter is to use $h$ hash functions to map the elements of a set onto an array of size $m$.  
To achieve this, each element is hashed $h$ times using different hash functions. The resulting hash values are used to index into the array, and the corresponding array cells are incremented by one. Since multiple elements can map to the same cell due to collisions, the counters of an element may not be incremented equally. To estimate the count of an element, the minimum value of the cells to which the element is mapped is used. The rationale behind this is that the minimum count is less likely to have been affected by collisions with other elements.

This is a highly efficient method for counting the number of packets per flow (or packets/flow) that have been collected in a memory block within a given time window. While a single hash function, such as CRC32, can also perform this task, counting Bloom filters are less prone to collisions, which can lead to inaccuracies in the count.

\subsection{Neural network architecture}\label{sec:lucid}

To validate our approach to data plane traffic feature extraction, we consider the \ac{ddos} attack detection use case. In particular, we focus on volumetric \ac{ddos} attacks, as they are particularly challenging to handle in constrained systems, such as network switches and SmartNICs, due to the large amounts of data rate such attacks can produce.

In this work, we adopt a state-of-the-art solution for \ac{ddos} attack detection called \ac{lucid} \cite{lucid}. \ac{lucid} is a \ac{cnn} that takes as input a representation of a traffic flow consisting of packet-level features and returns the probability of the flow being malicious (i.e., part of a \ac{ddos} attack). Its input format makes \ac{lucid} suitable for data plane feature extraction, where line-rate packet processing is a requirement. 

\ac{lucid}'s representation of traffic flows consists of packet-level features organised in two-dimensional arrays. Rows are the flow's packets in chronological order (\ac{lucid} defines bi-directional flows identified by a 5-tuple of IP addresses, L4 ports and L4 protocol), while columns are per-packet features, categorical features included (e.g., TCP flags, IP Flags, ICMP type, etc.). 
By utilising a convolutional layer as its initial hidden layer, \ac{lucid} effectively leverages the aforementioned representation to learn traffic flow semantics and uncover latent behavioural patterns from the chronological sequence of packets. \ac{lucid}'s output layer is a 1-neuron layer whose value is the probability of the input flow being a \ac{ddos} flow. 

We set the same hyper-parameters as in the \ac{lucid} paper with respect to the height and number of the convolutional kernels, $h=3$ and $k=64$ respectively. 
On the other hand, we slightly adapt the neural network's architecture to comply with the requirements of the feature extraction executed in the data plane. First, two packet features used by \ac{lucid}, namely \textit{highest layer} and \textit{protocols}, involve application layer information that is not available in the packet headers (\ac{lucid} extracts such features with the support of \textit{TShark} \cite{tshark}, which can return detailed packet's summaries along with standard header fields). Therefore, we only use the $f=9$ features that can be extracted in the data plane, namely: \textit{Timestamp}, \textit{Packet Length}, \textit{IP Flags}, \textit{TCP Length}, \textit{TCP Ack}, \textit{TCP Flags}, \textit{TCP Window Size}, \textit{UDP Length} and \textit{ICMP Type}.

We also decrease the detection time window from $t=100$ seconds to $t=2$ seconds and the number of packets/flow from $p=100$ to $p=4$. First, by reducing the input size we reduce the processing time and memory usage. Second, based on the results reported in \cite{lucid}, \ac{lucid} can achieve a very high classification accuracy with any value of $t>1$ second when combined with $p>3$. 

In summary, in our experiments, we use the following \ac{lucid} hyper-parameters:
$$\textbf{p}=4,\ \textbf{f}=9,\ \textbf{k}=64,\ \textbf{h}=3,\ \textbf{t}=2$$
Despite a lower number of packets and features in the flow representations and a shorter time window, we will show that the classification accuracy of the proposed system is comparable to that obtained by the original \ac{lucid} implementation on a publicly available dataset of \ac{ddos} and benign traffic.

\section{System Architecture}\label{sec:architecture}

Motivated by the insights presented in Section \ref{sec:introduction}, the objective of this study is to enable efficient network intrusion detection in the control plane of programmable networks through packet-level analysis. 
The primary challenge in achieving this objective lies in the limitations of the network devices responsible for extracting and collecting such features, as well as the constrained capacity of the control channel used to transmit them to the control plane. 

\begin{figure}[h!]
	\begin{center}
		\includegraphics[width=1\linewidth]{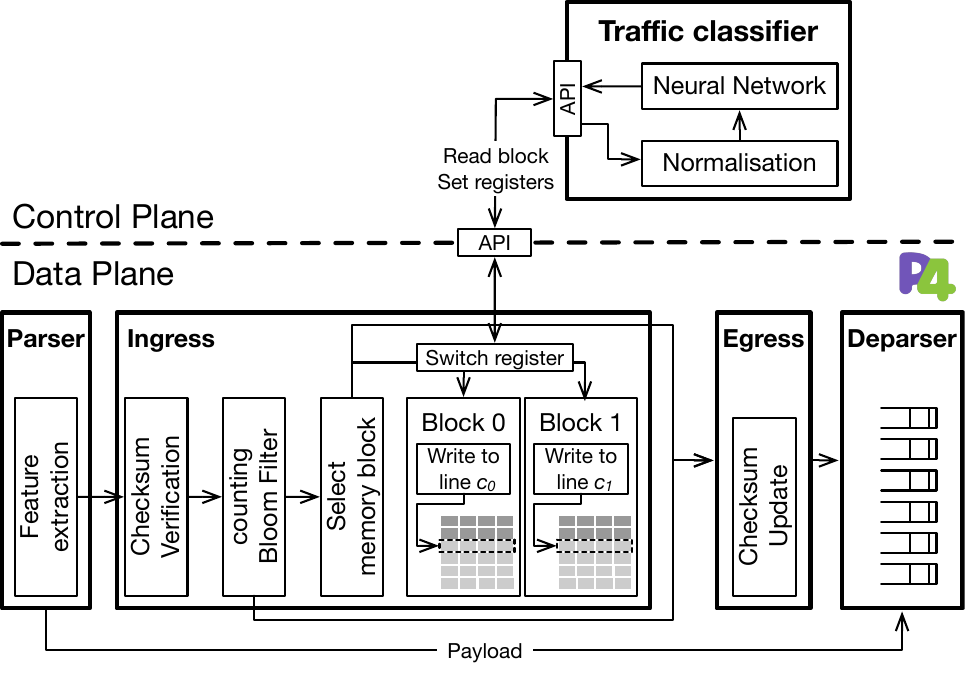}
		\caption{System architecture.}
		\label{fig:architecture}
	\end{center}
\end{figure}

Given these premises, we present \ac{ourtool}, an approach to intrusion detection for \ac{p4}-based programmable networks that enables resource-efficient traffic feature extraction and filtering in the data plane. The key idea of \ac{ourtool} is to extract and store in the data plane only the features that are essential to the \textit{Traffic Classifier} running on the control plane, thereby optimising the amount of data transmitted through the control channel.
For this purpose, we have designed a system architecture matching the \revised{\textit{V1model}} presented in Section \ref{sec:p4_programming}, with the support of control plane software for traffic classification (Figure \ref{fig:architecture}). The \ac{p4} program executed in the data plane extracts the packet attributes and makes them available to the control plane. On the other hand, the control plane is responsible for gathering the traffic metadata from the registers allocated in the data plane and coordinating the read/write access to such registers. 

The architecture presented in the figure has been specifically designed to reduce the workload on the control channel while adhering to the limitations imposed by the \ac{p4} language and the available resources in the data plane. 

\subsection{Core logic}
Data plane operations are sketched in Figure \ref{fig:architecture} (bottom part) and elaborated in the pseudo-code of Algorithm \ref{lst:feature-extraction}.
The data plane logic is implemented in the \textit{Parser} and in the \textit{Ingress Pipeline}. The \textit{Parser} takes the incoming packets and extracts header fields and metadata required by the \textit{Traffic Classifier} (line \ref{lst:feature-extraction:parser} of the pseudo-code). 
The components in the \textit{Ingress Pipeline} are in charge of storing the traffic features and other metadata in stateful memory, provided that the packet passes the  \textit{Checksum verification} (line \ref{lst:feature-extraction:verifychecksum}).

The key idea behind \ac{ourtool} for supporting packet-level feature extraction and storage is to collect only the features that are needed by the \textit{Traffic Classifier}. 
In our implementation, the traffic classifier (i.e., \textsc{Lucid}) takes as input a representation of traffic flows consisting of matrices of packet-level features (Section \ref{sec:lucid}). 
Each flow has the shape $p\times f$, where $p$ denotes the number of packets/flow, and $f$ represents the number of features extracted from the header of each packet. 
These requirements determine how the packet-level features are extracted and stored in the data plane. Nevertheless, it is important to note that our design is flexible and compatible with various packet-level representations of network traffic, including those involving segments of the packet's payload (e.g., FC-Net \cite{xu2020method}). 

\begin{algorithm}[t!]
	\caption{Data plane feature extraction and storage.}
	\label{lst:feature-extraction}
	\begin{algorithmic}[1]
		\renewcommand{\algorithmicrequire}{\textbf{Input:}}
		\Procedure{FeatureExtraction}{Packet ($pkt$)}
		\State \textbf{define} $p\times f$ \Comment Traffic Classifier input shape (Sec. \ref{sec:lucid}) \label{lst:feature-extraction:input-shape}
		\State $hdr,meta\gets$\Call{Parser}{$pkt$}\Comment Header and metadata\label{lst:feature-extraction:parser}
		\If {\Call{VerifyChecksum}{$hdr$}==\textbf{false}}\label{lst:feature-extraction:verifychecksum}
		\State \textbf{return}
		\EndIf
		\State $\bar{f}\gets meta.features(f)$ \Comment Vector of packet's features \label{lst:feature-extraction:readfeatures}
		\State $id_f^4,id_b^4,id_f^5,id_b^5\gets meta.id$\Comment Set of packet's IDs \label{lst:feature-extraction:packetid}
		\ForAll {$id\in \{id_f^4,id_b^4,id_f^5,id_b^5\}$}
		\State $h_{id}\gets\Call{CRC32}{id}\mod 2^r$ \Comment $h_{id}\in[0,2^r-1]$ \label{lst:feature-extraction:crc}
		\EndFor
		\State $k\gets$ \Call{SwitchReg.Read}{ } \Comment Memory index $k\in\{0,1\}$ \label{lst:feature-extraction:blockreg}
		\If {$min_{id\in \{id_f^4,id_b^4,id_f^5,id_b^5\}}F_k[h_{id}] < p$} \label{lst:feature-extraction:min-counter}
		\State $c_k\gets\Call{PositionReg.Read}{k}$ \label{lst:feature-extraction:position}
		\State \Call{Block.Write}{$B_k[c_k]$,$(id_f^5,\bar{f})$}\label{lst:feature-extraction:writefeatures}
		\State $c_k\gets c_k+1$
		\ForAll {$id\in \{id_f^4,id_b^4,id_f^5,id_b^5\}$}
		\If {$F_k[h_{id}] < p$}
		\State $F_k[h_{id}] += 1$ \Comment Increase the counters \label{lst:feature-extraction:increase-counters}
		\EndIf
		\EndFor
		\EndIf \label{lst:feature-extraction:end-min-counter}
		\State \Call{ComputeChecksum}{$h$} \label{lst:feature-extraction:computechecksum}
		\State \Call{DeParser}{$pkt$,$hdr$} \label{lst:feature-extraction:deparser}
		\EndProcedure
	\end{algorithmic}
\end{algorithm}

For each incoming packet, the $f$ features described in Section \ref{sec:lucid} are extracted from its header (line \ref{lst:feature-extraction:readfeatures}). Before storing this information in the stateful memory, \ac{ourtool} verifies that there are less than $p$ packets belonging to the same flow already stored in memory. If this condition is verified, the features are saved in memory, otherwise, they are discarded. By disregarding irrelevant features in the data plane, \ac{ourtool} ensures optimal utilisation of memory resources and control channel. 

To achieve this goal, \ac{ourtool} keeps track of the number of packets/flow collected within a given observation time window by using a probabilistic technique based on hashing called \textit{counting Bloom filters} \cite{tarkoma2011theory} described in Section \ref{sec:counting-bloom-filter}. Instead of using $h$ different hash functions to generate $h$ hash values from a packet, \ac{ourtool} leverages the \ac{p4} implementation of the CRC32 algorithm with $h=4$ different packet identifiers $id_f^4,id_b^4,id_f^5,id_b^5$ (line  \ref{lst:feature-extraction:packetid}).
The identifiers $id_f^4$ and $id_b^4$ each represent a 4-tuple consisting of source and destination IP addresses, as well as source and destination transport ports, for a given packet. The index $f$ refers to the \textit{forward} order, in which the values appear in the packet, while the index $b$ represents the \textit{backward} order, in which the positions of the IP addresses and transport ports are swapped. The other identifiers $id_f^5$ and $id_b^5$ are generated by taking the two 4-tuples $id_f^4$ and $id_b^4$, and by adding a fifth item with the value of the packet's transport protocol to each of them (i.e., they both are 5-tuples).

After computing the CRC32 value for each of the four identifiers (as shown in line \ref{lst:feature-extraction:crc}), the algorithm checks whether the minimum counter value among the four counters stored in the filter in the computed positions (i.e., hashed values) $F_k[h_{id}]$ is less than the maximum number of packets/flow allowed (i.e., $p$). If such a condition is verified, the current packet belongs to a network flow for which less than $p$ packets have been collected so far. In such a case, the packet's 5-tuple identifier $id_f^5$ and its corresponding features $f$ are saved in memory and the position in the memory is updated. These operations are summarised from line \ref{lst:feature-extraction:blockreg} to line \ref{lst:feature-extraction:end-min-counter} of the algorithm's pseudo-code and detailed in the following sections.

Finally, once the feature extraction and collection operations have been completed, the program updates the packet's checksum to reflect any packet modifications (even though the current implementation does not actually alter the packet's contents) (line \ref{lst:feature-extraction:computechecksum}). The header and payload are then reassembled through the \textsc{DeParser} function before the packet is sent to the egress port (line \ref{lst:feature-extraction:deparser}).

\subsection{Registers}\label{sec:registers}
Feature extraction and collection are handled with the support of stateful memory elements, such as registers. 
The value $k\in \{0,1\}$ of the \textit{Switch Register} is used to coordinate the memory access and to avoid race conditions between control and data planes. This value is set by the control plane and used by the data plane to select the appropriate registers for read/write operations (line \ref{lst:feature-extraction:blockreg}).

\begin{figure}[h!]
	\begin{center}
		\includegraphics[width=0.9\linewidth]{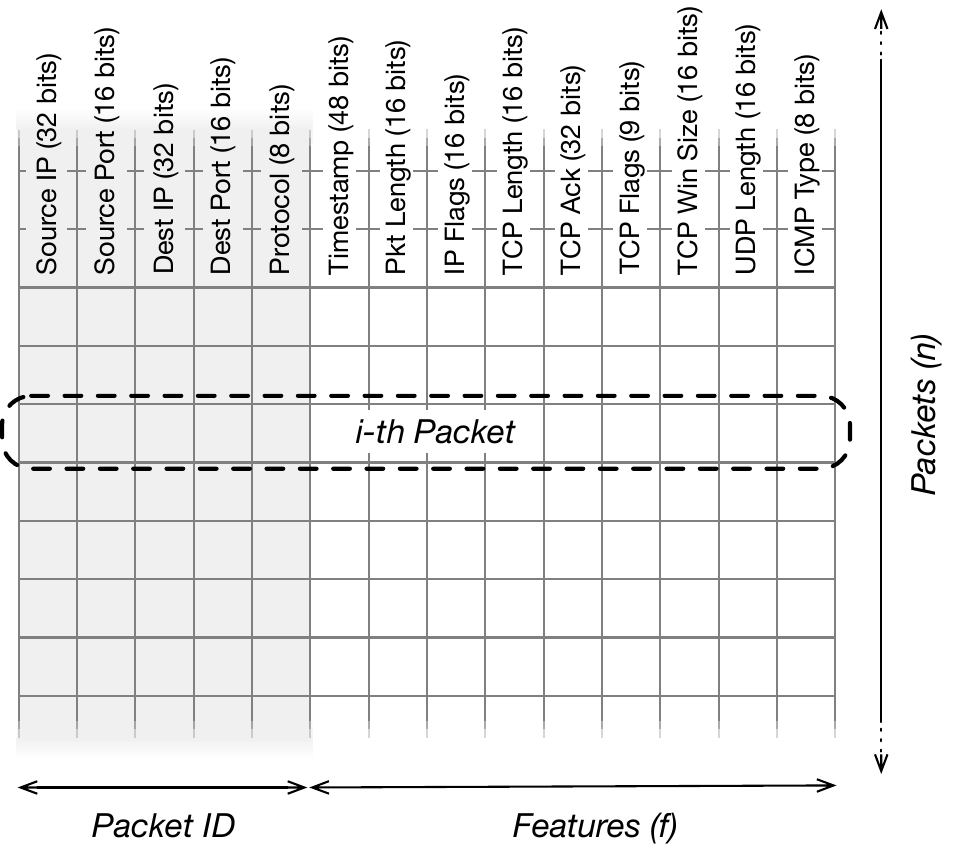}
		\caption{Memory block structure.}
		\label{fig:memory-block}
	\end{center}
\end{figure}

Traffic features are stored in two memory blocks $B_k$ (line \ref{lst:feature-extraction:writefeatures}). A memory block consists of $f$ registers (one per packet feature), along with other $5$ registers for hosting the five elements of the 5-tuple $id_f^5$, which is used later on by the \textit{Traffic Classifier} to map packets into flows. Each register is divided into $n$ cells, whose size varies depending on the value being stored. In our implementation, register \textit{Source IP} is of size  \texttt{<bit<32>>}, while \textit{Timestamp} is of size \texttt{<bit<48>>}, \textit{TCP Length} is of size \texttt{<bit<16>>}, etc. $n$ is the maximum number of packets whose features can be extracted and stored in $B_{k}$.
Figure \ref{fig:memory-block} sketches the structure of a memory block, in which each row represents a packet, and each column corresponds to an element of the 5-tuple packet identifier or a packet-level feature. 

In our implementation, we use two counting Bloom filters $F_k$, each one associated to the corresponding memory block $B_k$. Specifically, a filter is a register consisting of $m$ cells, each with a size of $3$ bits to store the packets/flow counters.
It is important to note that the output of the CRC32 hashing function is a 32-bit number, which would ideally require the allocation of two counting Bloom filters of $m=2^{32}$ cells. While this would be the optimal solution to minimize the chances of collisions, such an approach would require a significant amount of memory space. To address this concern, we have decided to limit the bit count of CRC32's output to $r<32$. We achieve this by using the \textit{modulo} operator (line \ref{lst:feature-extraction:crc}), with $r$ being a number dependent on the number of packets that can be stored in a memory block, denoted by $n$.  The sensitivity of \ac{ourtool} to the size of the counting Bloom Filter is provided in Section \ref{sec:bloom_filter_size}.
  
\subsection{Memory Management}\label{sec:memory_management}
The memory blocks $B_k$ are managed as circular buffers to ensure that the most recent traffic data is always available. Thus, when a memory block $B_k$ is full, denoted by $c_k=n-1$, the subsequent packet is stored at the beginning of the block, which corresponds to position $c_k=0$, overwriting the old packet stored there. When it happens for many packets, such as in high packet rate situations, the control plane classifier may miss the information of older flows either partially or entirely.

\subsection{Control-Data Plane interaction}
The Control Plane interacts with the Data Plane through \acp{rpc}, such as those provided by software frameworks like Apache Thrift API \cite{apachethrift} or P4Runtime API \cite{p4runtime}. In this regard, the Control Plane is in charge of swapping the value of the \textit{Switch Register}, collecting the packet features from the two memory blocks $B_k$ and clearing registers $B_k$, $F_k$ and $c_k$ ($k\in\{0,1\}$).
  
\begin{figure}[h!]
	\begin{center}
		\includegraphics[width=0.9\linewidth]{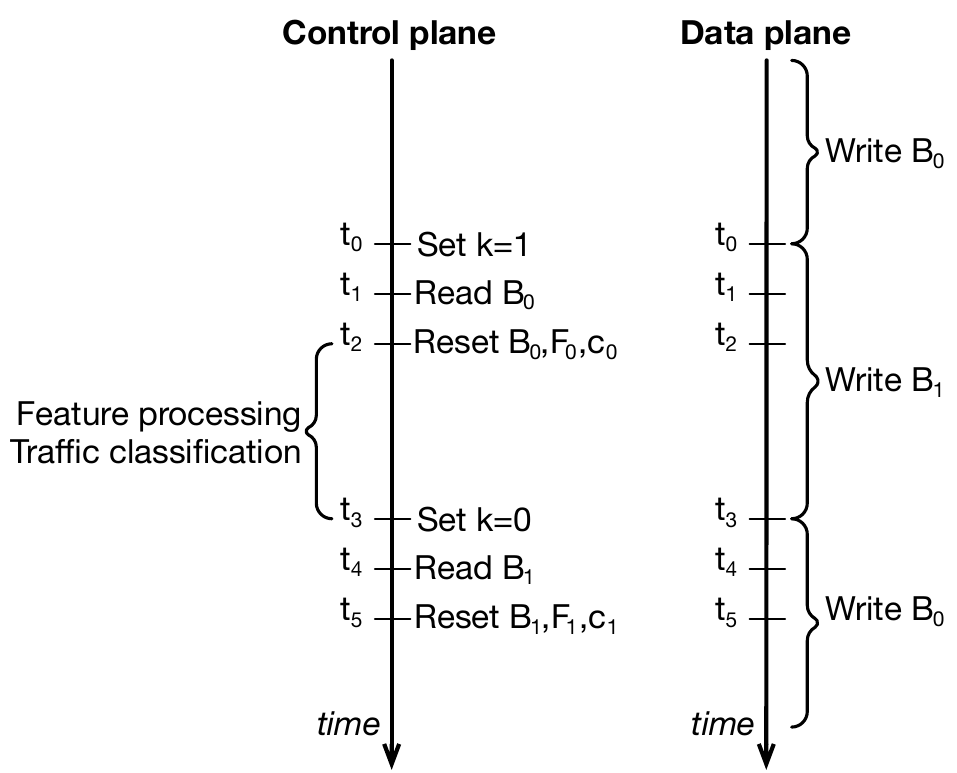}
		\caption{Memory access timeline.}
		\label{fig:memory-access}
	\end{center}
\end{figure}

Figure \ref{fig:memory-access} illustrates the timelines of Control and Data Plane operations. The two planes operate in parallel and never block each other. The Data Plane always writes on the same block until the value of the \textit{Switch Register} is changed from the Control Plane. If we start from time $t_0$, as shown in the figure, once the control plane sets the value of the \textit{Switch Register} to $1$, the Data Plane will start writing on memory block $B_1$. After that, the Control Plane will first read block $B_0$ and, once done with reading the block, it will reset all the registers of blocks $B_0$, $F_0$ and $c_0$ and the traffic features will be processed for classification. The whole process restarts at time $t_3$, when the Control Plane sets the \textit{Switch Register} to $0$ to retrieve the features collected in memory block $B_1$, while the Data Plane switches to block $B_0$. 
\revised{It is important to note that the frequency of this process (computed as $1/(t_3-t_0)$ using the example in Figure \ref{fig:memory-access}) is managed by the Control Plane and does not depend on Data Plane's state.}
\section{Experimental setup}\label{sec:setup}

\ac{ourtool} has been implemented as a set of methods for the reference P4 software switch (namely the \ac{bmv2} \cite{bmv2}) and tested in a Mininet emulated network \cite{mininet}. A prototype implementation of \ac{ourtool} is publicly available for testing and use \cite{p4ddle-source}. 

\begin{figure}[h!]
	\begin{center}
		\includegraphics[width=0.8\linewidth]{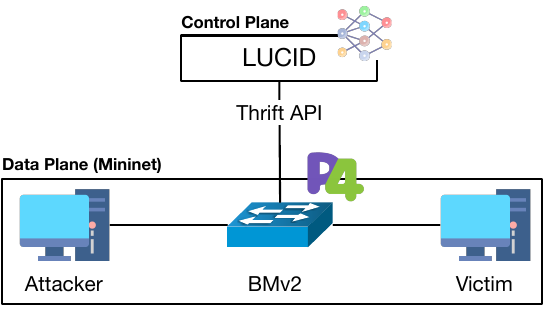}
		\caption{Experimental setup.}
		\label{fig:testbed}
	\end{center}
\end{figure}

The network topology, represented in Figure \ref{fig:testbed}, includes two virtual hosts, one acting as the attacker which sends malicious traffic to the second virtual host (i.e., the victim of the attack).  The evaluation environment has been set up on a single physical machine equipped with an 80-core Intel(R) Xeon(R) Gold 5218R CPU @2.10GHz and 128 GB of DDR4 RAM. This machine also runs the \ac{lucid} framework \cite{lucid-github}, which includes a pre-processing tool and a \ac{cnn} trained for \ac{ddos} attack detection. 

\ac{lucid} and the switch are interfaced through \acfp{rpc}. The \ac{rpc} mechanism is implemented using the Apache Thrift API \cite{apachethrift} and is used to read, write and reset the registers from the control plane through data plane methods \texttt{bm\_register\_read}, \texttt{bm\_register\_write} and \texttt{bm\_register\_reset} respectively \cite{bmthrift}.  

\subsection{Dataset\label{sec:dataset}}
\ac{ourtool} is evaluated using CIC-DDoS2019 \cite{sharafaldin2019developing}, a recent dataset of \ac{ddos} attacks provided by the Canadian Institute of Cybersecurity at the University of New Brunswick. This dataset contains multiple days of network activity, including benign traffic and 13 distinct types of \ac{ddos} attacks. It is accessible to the public and contains pre-recorded traffic traces with complete packet payloads, along with supplementary text files that provide labels and statistical information for each traffic flow \cite{cicddos2019}. In our experiments, we inject the \ac{p4} switch with the traffic traces to evaluate \ac{ourtool}'s performance in handling the feature extraction process in the data plane.
In the dataset, the benign traffic was created using the B-profile \cite{sharafaldin2018towards}, which defines distribution models for various applications, such as web (HTTP/S), remote shell (SSH), file transfer (FTP), and email (SMTP). On the other hand, the attack traffic was generated using third-party tools and can be broadly classified into two main categories: \textit{reflection-based} and \textit{exploitation-based} attacks. The first category involves attacks in which the attacker triggers responses from a remote server (such as a DNS resolver) towards the victim's spoofed IP address, ultimately overwhelming the victim with these responses. 
The second category pertains to attacks that exploit known vulnerabilities in target systems, applications or in certain network protocols.

The traffic traces have been divided into training, validation and test sets and processed with the \ac{lucid}'s packet parser \cite{lucid-parser}. Such a tool extracts packet-level features from the traffic and groups them into array-like representations of network flows, as described in Section \ref{sec:lucid}. 

\begin{table}[h]
	\caption{\ac{ddos} attack traces (test sets).} 
	\label{tab:dataset}
	\small 
	\centering 
	\begin{tabular}{lcccc} \toprule[\heavyrulewidth]
		\multirow{2}{*}{{\textbf{Attack}}}  
		& \textbf{Duration} &\textbf{\#Flows} &\textbf{Flow rate} &\textbf{Packet rate}  \\ 
		& \textbf{(sec)} & &\textbf{(flows/sec)} &\textbf{(kpackets/sec)}  \\  \midrule 
		\textbf{DNS}  & 383 & 5736 & 15 & 2.6\\ 
		\textbf{LDAP}   & 27 & 26 & 0.96 &  8 \\ 
		\textbf{MSSQL}   & 65 & 787092 & 12109 & 24 \\ 
		\textbf{NetBIOS}   & 65 & 519069 & 7985 &  16\\ 
		\textbf{Portmap} & 19 & 152956 & 8050 & 16 \\  
		\textbf{SNMP}   & 38 & 74882 & 1970 &  17  \\  
		\textbf{SSDP}   & 23 & 195521 & 8500 & 30  \\  
		\textbf{UDP-Lag}  & 13 & 116439 & 8957 & 20   \\  	\bottomrule[\heavyrulewidth] 
	\end{tabular}
\end{table}

The pre-processing of the training and validation traces is done offline for \ac{lucid}'s training and tuning, whereas the traffic traces of the test set are solely employed for online testing.  During the testing phase, feature extraction is performed in the data plane (as explained in Section \ref{sec:architecture}), while \ac{lucid}'s parser tool is executed in the control plane for building the arrays and normalising the features. 

To mitigate the impact of \ac{bmv2}'s poor performance \cite{bmv2-performance}, we made a strategic decision to exclude all attacks with a packet rate higher than 30 kpackets/s from the test set. This step was necessary to ensure the integrity and accuracy of our evaluation results by preventing any interference caused by packets being dropped due to \ac{bmv2}'s performance limitations.
The key details of the remaining test traces are presented in Table \ref{tab:dataset} for reference.

\subsection{Evaluation metrics}\label{sec:metrics}
Our primary evaluation metrics are the \textit{Collected Flows} and the \textit{\acf{sfnr}}. 
\textit{Collected Flows} measures the number of flows stored in memory relative to the total number of flows injected into the switch within a given time window. This metric provides insight into \ac{ourtool}'s ability to capture information on as many traffic flows as possible. 

The \ac{sfnr} quantifies the percentage of malicious flows that go undetected, either due to misclassification by the \ac{nids}, or because no packets of such flows are present in the memory block due to the reasons explained in Section \ref{sec:memory_management}. By assessing the \ac{sfnr}, we measure the efficiency of \ac{ourtool} in promptly identifying and flagging potential intrusions. A formal definition of the \ac{sfnr} measured in a given observation time window is provided in Equation \ref{eq:sys_fnr}.

\begin{equation}\label{eq:sys_fnr}
	sFNR = FNR+\frac{1}{|X_m|}\sum_{x_i\in X_m} c_{x_i}\oplus n_{x_i} 
\end{equation}

In Equation \ref{eq:sys_fnr}, \ac{fnr} is the False Negative Rate of the \ac{nids} running in the control plane. It is worth reminding that \ac{fnr} is the metric that measures the percentage of positive samples (in our case, the malicious flows) that are misclassified as negatives by a classifier. 
In the equation, the set of malicious flows is denoted by $X_m$, $c_{x_i}$ represents the actual number of packets collected for a given malicious flow $x_i$ in a memory block, while $n_{x_i}$ is the number of observed packets of that flow. The \textit{XOR} operation $c_{x_i}\oplus n_{x_i}$ returns $1$ if $c_{x_i}=0$ and $n_{x_i}\neq 0$ or if $c_{x_i}\neq0$ and $n_{x_i}=0$ (the second case can never happen), and returns $0$ otherwise. The summation computes the total number of observed malicious flows in a given time interval with no packets collected in the memory block. We divide this value by the total number of observed malicious flows $X_m$ to obtain the rate. 

We also define the \textit{quality} metric that allows us to establish a relationship between three distinct quantities: (i) the number of packets in a traffic flow, (ii) the number of packets of that flow collected in the data plane within a given time window, and (iii) the packet/flow $p$ required by the \textit{Traffic Classifier} running in the control plane. This metric quantifies the level of ``useful'' information that the data plane delivers to the Traffic Classifier for each flow, while taking into account that gathering more than $p$ packets per flow is inefficient resource utilization. Specifically, the quality metric is defined by Equation \ref{eq:quality}. 

\begin{equation}\label{eq:quality}
quality = \frac{1}{|X|}\sum_{x_i\in X} \frac{c_{x_i}}{l_{x_i}}
\end{equation}

The quality metric is determined by taking the average quality score across the flows observed within a given time window. In the equation, the set of such flows is denoted by $X=\{X_b,X_m\}$ and includes both benign flows $X_b$ and malicious flows $X_m$. The quality of a single flow $x_i\in X$ is expressed as $c_{x_i}$ over $l_{x_i}$, where $l_{x_i} = \min \{n_{x_i}, p\}$ is the minimum between the number of packets $n_{x_i}$ of the $i$-th flow and $p$. $l_{x_i}$ represents the optimal number of packets that we need to collect in the data plane to maximise the amount of information provided to the \ac{ml} model for a given flow $x_i$.
Collecting fewer packets may not provide sufficient information for the classifier, while collecting more may waste valuable memory space without providing any additional benefits to the classifier. This is because any excess packets beyond the optimal number would be discarded by the feature pre-processing component, which constructs the flow samples required by the \ac{ml} model.

In the case of the quality metric, $c_{x_i}$ represents the number of packets stored in a memory block for a given flow $x_i$, either benign or malicious. With \ac{ourtool}, which uses a counting Bloom filter to track the number of packets/flow in memory, the value of $c_{x_i}$ ranges from $0$ to $p$. With no packet tracking,  $c_{x_i}$ may fall between $0$ and $n_{x_i}$.
It is worth noting that in certain cases, the value of $c_{x_i}$ may be zero even when $n_{x_i}\neq 0$. This can occur when no packets for flow $x_i$ have been collected in the current memory block, either because of collisions or because they were overwritten by more recent packets in the circular buffer (Section \ref{sec:memory_management}).
\section{Evaluation Results}\label{sec:evaluation}
One of the key benefits of \ac{ourtool} over other approaches is the ability to extract raw packet features from the network traffic, categorical features included, and to organise them in a way that the semantics of traffic flows are preserved. 
\ac{ourtool} efficiently achieves this objective by implementing a counting Bloom filter (Section \ref{sec:counting-bloom-filter}) that tracks the number of packets/flow without any wastage of memory resources in the data plane. 
We demonstrate the advantages of \ac{ourtool} through simulation and emulation experiments, where we disable the tracking methods and observe the corresponding changes using a range of metrics.

In the simulation scenario, no actual network data is involved. Instead, we rely on the network profile of a campus network \cite{jurkiewicz2021} and we demonstrate the benefits of \ac{ourtool} by simulating the feature extraction and storage process in the data plane.  
The second set of experiments has been conducted in an emulated environment consisting of two hosts (an attacker and a victim) and a \ac{p4}-enabled software switch with configurable registers. The experiments involve injecting pre-recorded network traffic into this environment, using a publicly available dataset of both benign and \ac{ddos} traffic.

\subsection{Simulation Scenario}

The goal of the simulations is to analyse the correlation between the size of the memory block $B_k$ and two key variables: the maximum number of packets/flow (i.e., variable $p$) and the number of cells $m$ of the Bloom filter $F_k$. This information gives an understanding on how to configure \ac{ourtool} based on the characteristics of the network traffic under monitoring (mainly the average packet rate and average packets/flow) to maximise the number of collected flows within a given time window. To do so, we fix the size of $B_k$ and we vary the value of $p$ and the bit count $r$ of the CRC32's output that determines the size of the Bloom filter (see Section \ref{sec:registers}).

We utilise the profile outlined in \cite{jurkiewicz2021} to simulate the \revised{benign} traffic of a realistic network. This profile was generated using network activity data gathered from a university campus \revised{and does not present any documented attack in the trace}. From it, we extracted the distribution of TCP, UDP, and other protocols and we computed the average number of packets/flow for each protocol.

Each experiment consists of $1000$ iterations, each simulating the collection of a random number of flows, ranging from 1 to 8192 flows. The length of such flows (number of packets/flow) is generated based on the aforementioned traffic profiles, while the arrival of packets across different flows has been randomised to ensure a non-sequential packet distribution\revised{, aligning with the characteristics of realistic packet-switched networks.}
To minimise potential issues due to the insufficient space in memory to store packets for all the 8192 flows (which is not the goal of this analysis), we reserve space in the memory block $B_k$ for at least two packets/flow by setting the value $n=16384$ packets (see also Figure \ref{fig:memory-block} for reference).

\revised{
These experiments are realised with Python scripts designed to simulate a single collection of network flows. The scripts replicate the logic of both \ac{ourtool} (as described in Algorithm \ref{lst:feature-extraction}) and a baseline configuration. 
In the baseline setup, the packet filtering algorithm intrinsic to \ac{ourtool} is deliberately deactivated, causing all incoming packets to be stored in memory within the circular buffer, with no constraints on the number of packets per flow.
These scripts are publicly available for testing on the \ac{ourtool} repository \cite{p4ddle-source}.}

\subsubsection{Bloom filter size}\label{sec:bloom_filter_size}

Each Bloom filter is characterised by two key dimensions: the number of cells, denoted as $m\in [0,2^r-1]$, and the size of each cell. In this experiment, we have set the size of the cells while varying the value of $r$ to understand the sensitivity of \ac{ourtool} to the number of cells in terms of \textit{Collected Flows}.
Considering that 75\% of the flows within the traffic profile consist of no more than 4 packets, we have chosen to set the cell size to 3 bits. This size adequately accommodates the counting of up to 4 packets/flow.

\begin{figure}[h!]
	\begin{center}
		\includegraphics[width=\linewidth]{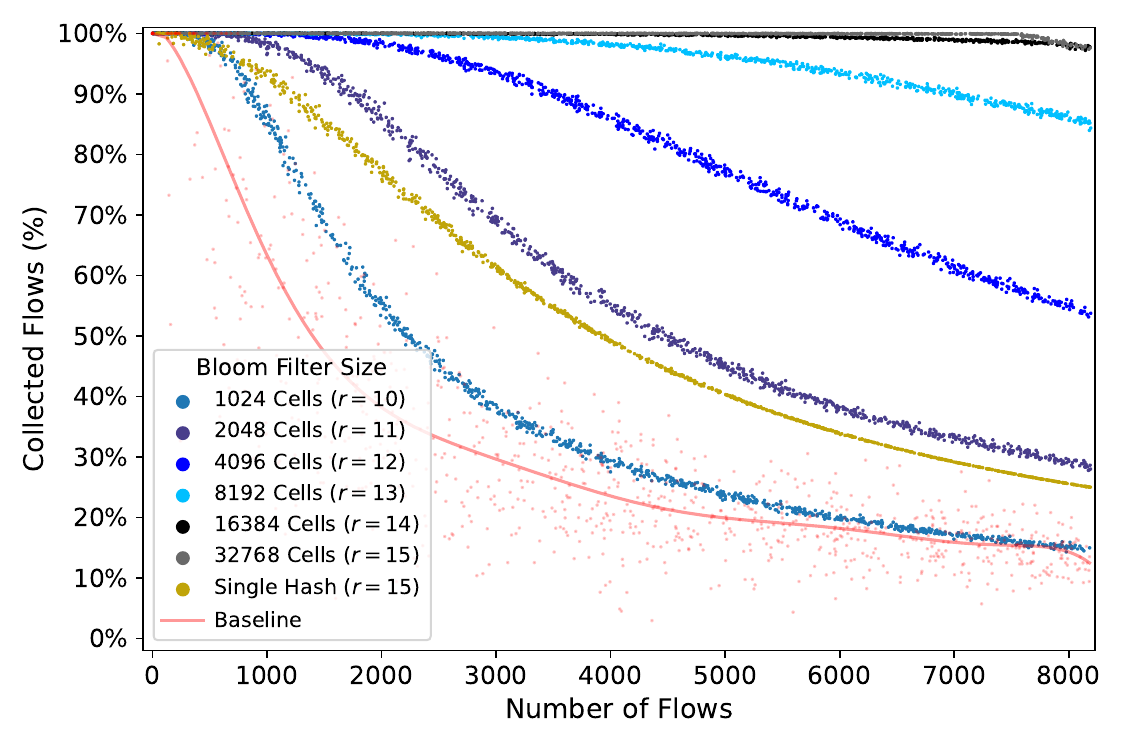}
		\caption{Percentage of collected flows on the value of $r$.}
		\label{fig:collected_flows_per_bf_size}
	\end{center}
\end{figure}

Figure \ref{fig:collected_flows_per_bf_size} presents the results obtained with Bloom filter sizes from $1024$ to $32768$ cells (or $r$ from $10$ to $15$). \revised{The figure also includes the results obtained with a single hash function (with $r=15$) , allowing for a comparative analysis with Bloom filters.}
As expected, the larger the Bloom filter size, the smaller the number of hash collisions and the higher the number of flows collected. 
Remarkably, it is worth noting that \ac{ourtool} is able to collect more flows than the baseline, regardless of the Bloom filter size. It is important to recall that the baseline approach gathers all packets without any filtering logic. Considering that the average number of packets/flow in the traffic profile is 78, the baseline approach exhausts the memory capacity with approximately $210$ flows (calculated as $16384/78$). Consequently, once the memory block reaches its limit, older packets are overwritten by more recent ones, resulting in a loss of flows.
\revised{The results of this experiment also confirm the efficiency of Bloom filters over a single hash function in limiting the collisions, as discussed in Section \ref{sec:counting-bloom-filter} and demonstrated by Tarkoma et al. \cite{tarkoma2011theory}}.

Figure \ref{fig:collected_flows_per_bf_size} leads to another significant observation: the stability of \ac{ourtool} throughout the experiment rounds, in contrast to the baseline approach. This stability is a direct result of the packet filtering strategy employed by \ac{ourtool}. By filtering out excess packets from each flow, \ac{ourtool} ensures that the memory allocated for a flow remains bounded by the predetermined maximum $p$. Consequently, the number of collected flows remains stable even in extremely randomised scenarios like this simulation. 
In contrast, the baseline approach lacks any form of packet filtering. As a result, the number of packets per flow stored in memory is not constrained, leading to much larger fluctuations in terms of collected flows. This absence of bounds on per-flow packet storage is evident from the erratic behaviour depicted in the figure, where the scattered values of collected flows reflect the wide range of flow sizes, spanning from 1 to 1000 packets (with average and standard deviation of 77 and 74 packets, respectively). The baseline curve shown in the figure represents a polynomial approximation of the values and serves to emphasize the average trend of collected flows under this configuration.

In light of the results obtained in this simulation, we have determined that setting the number of cells to $32768$ is the most suitable choice. 
This value effectively minimises hash collisions (0.5\% with 8000 flows) while maintaining a minimal impact on memory usage. It is important to note that \ac{ourtool} utilises two counting Bloom filters ($F_k$ with $k\in\{0,1\}$ in Algorithm \ref{lst:feature-extraction}).
Since we only need 3 bits to count up to $p=4$ packets/flow, the size of each $F_k$ Bloom filter will be 12 kilobytes. This memory requirement is relatively insignificant when compared to the memory blocks $B_k$, each of which demands approximately $562$ kilobytes to store the features of 16384 packets.

\subsubsection{Maximum number of collected packets per flow}

In this experiment, we fix the size of the Bloom filter to $32768$ cells, and we vary $p$ from $2$ to \revised{$128$} packets/flow. We expect that by decreasing the value of $p$, the average number of packets collected per flow will also decrease, leading to a higher number of collected flows in memory but also to more potential collisions.
\begin{figure}[h!]
	\begin{center}
		\includegraphics[width=\linewidth]{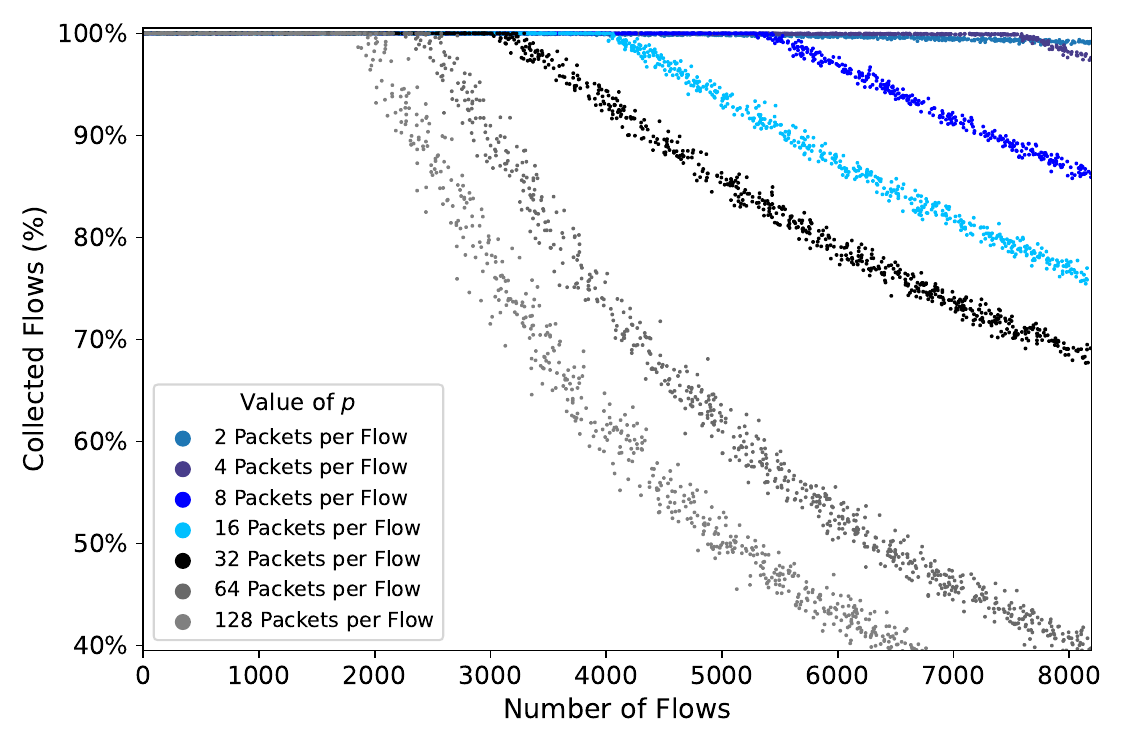}
		\caption{\revised{Percentage of collected flows on the value of $p$.}}
		\label{fig:quality_per_num_packets_per_flow}
	\end{center}
\end{figure}

This behaviour can be observed in Figure \ref{fig:quality_per_num_packets_per_flow}, which demonstrates that by increasing $p$, the memory block fills up sooner and, as a consequence, the percentage of collected flows deteriorates earlier. 
Of course, with $p=2$ the memory is never filled, even with $8192$ generated flows (we remind that we set the size of the memory block to $16384$ packets). However, we can notice that with $p=2$ a small portion of flows is lost due to collisions when the number of generated flows is $5000$ or higher. 
Indeed, when the value of $p$ is small, there is an increased likelihood of missing the condition at line \ref{lst:feature-extraction:min-counter} of Algorithm \ref{lst:feature-extraction}. As a result, a higher number of packets and flows are discarded.
In the figure, we can also notice the stability of \ac{ourtool} with $p=4$ almost until the end of the experiment. For this reason, and considering the proven effectiveness of \ac{lucid} with $p>3$ (as discussed in Section \ref{sec:lucid}), we set $p=4$ for the experiments conducted in the emulated environment.

\subsection{Emulated Scenario}\label{sec:emulated_scenario}

Based on the results of the simulations described in the previous section, we set the parameters for the emulations as follows: (i) memory block size: $n=16384$ packets, (ii) bloom filter size: $32768$ cells ($r=15$) and (iii) maximum packets/flows $p=4$. The duration of all the experiments has been set to $6$ minutes, during which we inject various combinations of attack and benign traffic from the attacking host to the victim through the \ac{ourtool}-enabled switch (see Figure \ref{fig:testbed}). 

In this experiment, we generate a series of \ac{ddos} attacks by using the traffic traces of the CIC-DDoS2019 dataset (see Section \ref{sec:dataset}). If we exclude the DNS attack, the flows of the other attacks present an average flow length of around $2.2$ packets, hence considerably smaller than those of the profile of the campus network used in simulations (around $78$ packets/flow). 
Considering that we chose $p=4$ (i.e., $p>2.2$), we expect that with these small flows, both \ac{ourtool} and the baseline can collect approximately the same number of flows. 
Acknowledging that malicious traffic flows typically do not constitute the totality of network traffic, we define four evaluation scenarios to enable a comprehensive comparison. These scenarios involve the inclusion of background benign traffic at varying proportions relative to the total packets in the network, namely 0\% (no benign traffic), 25\%, 50\%, and 75\%. Table \ref{table:avg_p_p_f} presents the average flow length as we increase the percentage of benign traffic in the network (\revised{whose average packet rate is about 0.7 packets/sec and} average flow length is about $36$ packets).

\begin{table}[t!]
	\centering
	\caption{\label{table:avg_p_p_f}Impact of benign traffic on the average flow length.}
	\begin{tabular}{lcccccccc}
	\toprule
	\multirow{2}{*}{\small{\textbf{Attack}}} & \multicolumn{4}{c}{\small{\textbf{Average Flow Length}}}                                                                                          \\
									  & \textbf{0\% } & \textbf{25\% }  & \textbf{50\% }  & \textbf{75\% }  \\
	\midrule
				 SSDP &   3.5 &  4.4 &  6.0 &  9.7 \\
				MSSQL &   2.0 &  2.7 &  3.8 &  6.8 \\
			  UDP-Lag &   2.2 &  2.9 &  4.1 &  7.3 \\
				 SNMP &   2.0 &  2.6 &  3.8 &  6.7 \\
			  Portmap &   2.0 &  2.6 &  3.7 &  6.6 \\
			  NetBIOS &   2.0 &  2.6 &  3.7 &  6.6 \\
				 LDAP &   2.0 &  2.6 &  3.7 &  6.6 \\
				  DNS & 109.3 & 76.3 & 66.0 & 49.8 \\
	\midrule
	
			  \textbf{Average} &  \textbf{15.6} & \textbf{12.1} & \textbf{11.9} & \textbf{12.5} \\
	\textbf{Average (w/o DNS)} &   \textbf{2.2} &  \textbf{2.9} &  \textbf{4.1} & \textbf{7.2} \\
	\bottomrule
	\end{tabular}
\end{table}

\begin{table}[t!]
\centering
\caption{\label{table:ddos_flows}Percentage of DDoS Flows based on the volume of benign packets.}
\begin{tabular}{lccccccccc}
\toprule
\multirow{2}{*}{\small{\textbf{Attack}}} & \multicolumn{4}{c}{\small{\textbf{Percentage of DDoS Flows}}}                                                                                          \\
& \textbf{0\%} & \textbf{25\%}  & \textbf{50\%}  & \textbf{75\%}  \\
\midrule
   SSDP & 100\% & 96.01\% & 90.07\% & 76.78\% \\
  MSSQL & 100\% & 97.11\% & 93.00\% & 83.40\% \\
UDP-Lag & 100\% & 97.40\% & 93.19\% & 82.50\% \\
   SNMP & 100\% & 97.63\% & 93.72\% & 83.45\% \\
Portmap & 100\% & 97.70\% & 93.57\% & 83.16\% \\
NetBIOS & 100\% & 97.71\% & 93.51\% & 83.20\% \\
   LDAP & 100\% & 96.99\% & 93.48\% & 83.16\% \\
    DNS & 100\% & 46.89\% & 27.15\% & 14.87\% \\
\midrule
\textbf{Average} & \textbf{100\%} & \textbf{90.93\%} & \textbf{84.71\%} & \textbf{73.82\%} \\
\bottomrule
\end{tabular}
\end{table}

Other important information pertains to the relationship between packets and flows. Since we add a portion of background benign traffic based on the total amount of packets, this does not necessarily reflect an analogous increase in the number of flows. Actually, based on the low average flow length in the attacks (first column in Table \ref{table:avg_p_p_f}), and the relatively high number of packets/flow of the benign trace (around $36$ on average), even splitting 50\% between benign and \ac{ddos} packets, the average percentage of \ac{ddos} flows is higher than 93\% on the total of flows (excluding the DNS attack that presents an exceptional behaviour). Table \ref{table:ddos_flows} shows the percentage of \ac{ddos} flows for each attack type.

Given these premises, we compare \ac{ourtool} against the baseline approach in the emulated environment. For both approaches, we inject each of the eight attacks of the CIC-DDoS2019 dataset combined with benign traffic in the various proportions reported above. The comparison is evaluated using the three metrics presented in Section \ref{sec:metrics}, namely \textit{\ac{sfnr}}, \textit{Collected Flows} and \textit{quality}.

\subsubsection{Collected Flows and sFNR}

\begin{figure*}[h!]
	\begin{center}
		\includegraphics[width=\linewidth]{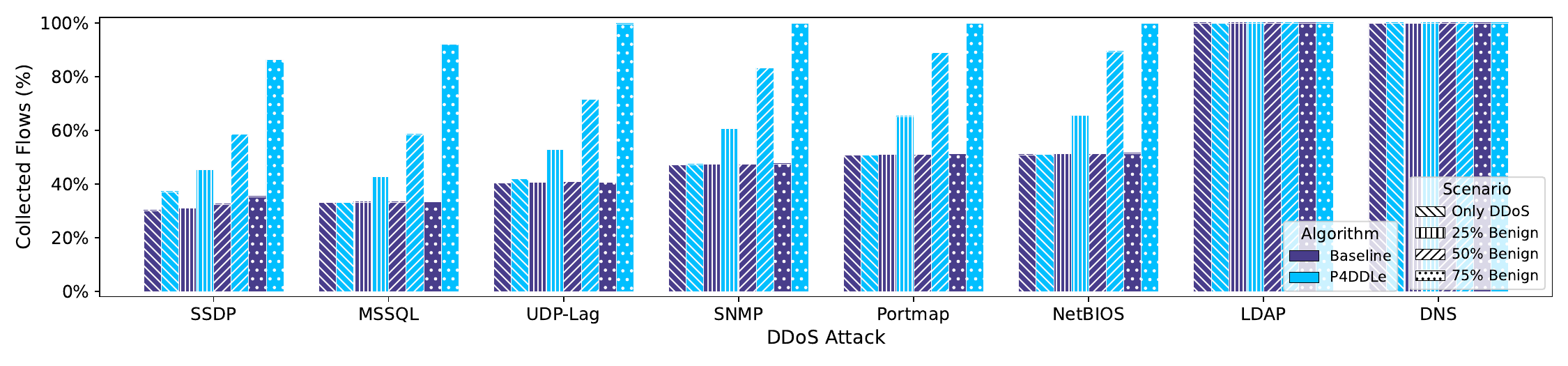}
		\caption{Average Collected Flows.}
		\label{fig:metrics-collected}
	\end{center}
\end{figure*}

\begin{figure*}[h!]
	\begin{center}
		\includegraphics[width=\linewidth]{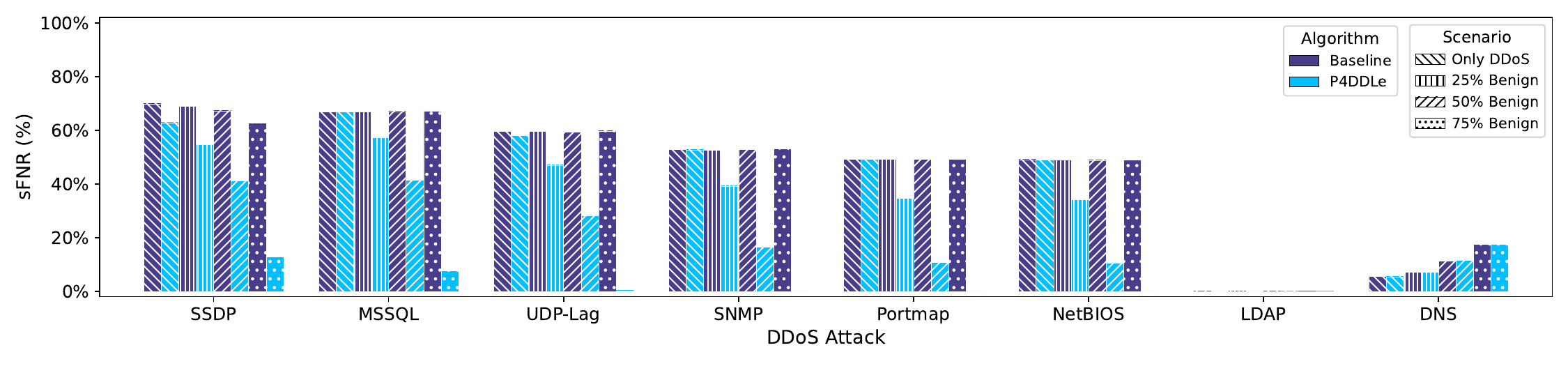}
		\caption{Average \ac{sfnr}.}
		\label{fig:metrics-sFNR}
	\end{center}
\end{figure*}

Figure \ref{fig:metrics-collected} shows the comparison related to \textit{Collected Flows} for all attack types. First, we can notice that with the baseline approach the percentage of collected flows remains stable when we vary the proportion between benign and attack traffic. This can be attributed to the absence of any filtering logic, causing the percentage of collected flows to solely depend on the memory capacity.  
In contrast, \ac{ourtool} demonstrates consistent improvement in the percentage of collected flows as the volume of benign traffic (and the average flow length) increases. While the baseline approach fills up the memory block predominantly with large benign flows, \ac{ourtool} optimizes memory usage by storing only the essential information required by the \ac{nids} and filtering out packets that exceed the maximum limit of $p=4$. This memory management prevents large flows from occupying excessive memory space, thus ensuring that a higher number of flows can be transmitted to the \ac{nids} for classification.
For instance, in the case of the MSSQL attack, whose average flow size is around $2$ packets/flow, with 0\% of benign traffic the performance of \ac{ourtool} is approximately the same as with the baseline approach. This is because with $p=4$, no packets are filtered and the memory occupation is similar with both approaches. However, as soon as we add some benign traffic, which is composed of larger flows, the filtering mechanism starts dropping the extra packets, saving space in memory for more flows. 

\revised{\ac{ourtool} inherently maintains a balanced allocation of memory for both low-rate and high-rate flows, resulting in similar probabilities for their storage. 
This balance can be observed when analyzing the distribution of collected flows, distinguishing between \ac{ddos} traffic, whose packet rate ranges from 2.6K to 30K packets per second, and benign flows, which average around 0.7 packets per second.
For instance, consider the SSDP attack scenario with $30$K packets/sec, where the test traffic is divided into $25\%$ benign and $75\%$ attack traffic. In this worst-case scenario, the average percentage of collected flows is approximately $42\%$ for benign flows and $46\%$ for attack flows, demonstrating \ac{ourtool}'s stability and effectiveness in handling heterogeneous network conditions.} 

The \ac{sfnr} metric (Figure \ref{fig:metrics-sFNR}) shows the ability of \ac{ourtool} to collect more malicious flows than the baseline approach, ultimately leading to faster mitigation of network attacks.
It is also worth noticing that when the volume of the attack is small, like in the case of LDAP and DNS, the memory is never filled, even when adding benign traffic.  Therefore, the collected flows is 100\% for both, while the non-zero \ac{sfnr} obtained with the DNS attacks is solely due to  \ac{lucid}'s \ac{fnr}.

\subsubsection{Flow quality}
As defined in Section \ref{sec:metrics}, the quality metric measures the average amount of ``useful'' data collected in the data plane and transmitted to the control plane for classification. Table \ref{tab:quality} presents the average quality metrics, revealing that while \ac{ourtool} can collect a larger number of flows in memory, these flows maintain a high level of quality, although sometimes slightly lower than that obtained with the baseline. 

\begin{figure*}[h!]
	\begin{center}
		\includegraphics[width=\linewidth]{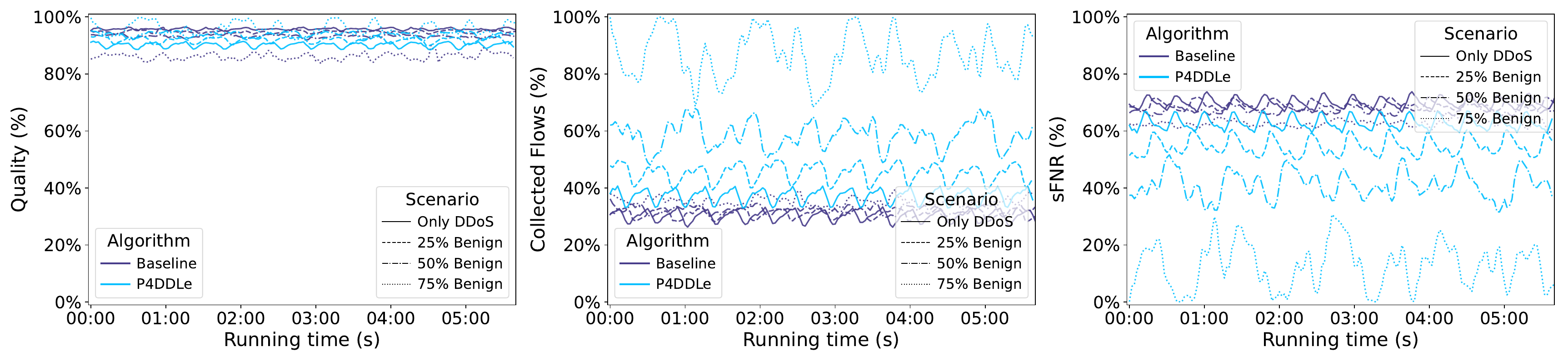}
		\caption{Metrics by running time on SSDP DDoS Attack.}
		\label{fig:metrics-ssdp}
	\end{center}
\end{figure*}
\begin{figure*}[h!]
	\begin{center}
		\includegraphics[width=\linewidth]{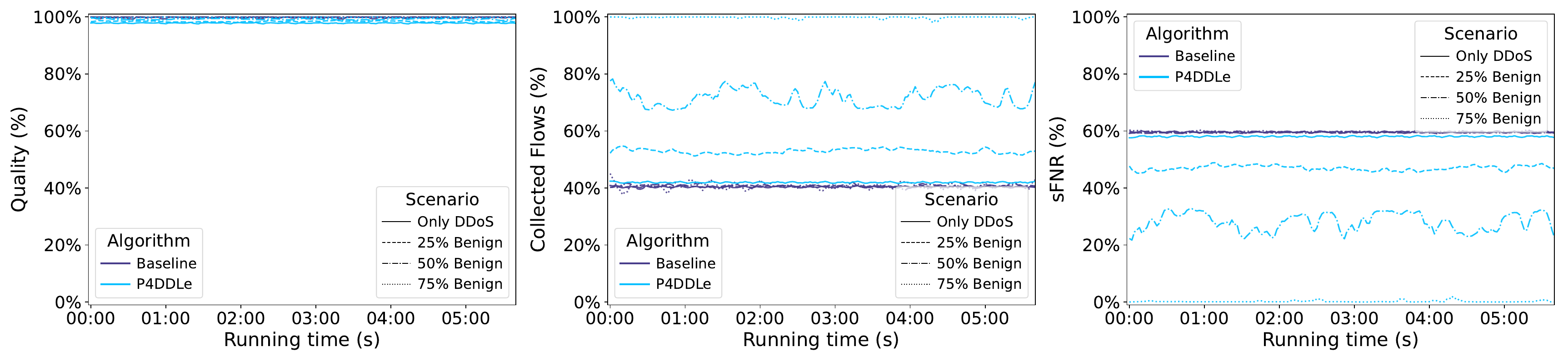}
		\caption{Metrics by running time on UDP-Lag DDoS Attack.}
		\label{fig:metrics-udplag}
	\end{center}
\end{figure*}

Unlike the baseline, \ac{ourtool} might drop packets because of collisions, which accounts for the slight reduction in quality observed in the attack scenarios that present a combination of high flow rate and high packet rate. The SSDP attack is an interesting use case that demonstrates this observation: as we add more and more benign traffic, the packet rate remains constant (around 30 kpackets/sec), while the flow rate decreases, hence the chances of collisions. 

\begin{table}[h!]
\centering
\caption{\label{tab:quality} Quality of collected flows at various percentages of benign packets.}
\begin{adjustbox}{width=0.49\textwidth}
\begin{tabular}{lcccccccc}
\toprule
\multirow{2}{*}{\small{\textbf{Attack}}}                                                                                        
                                 & \multicolumn{4}{c}{\small{\textbf{Baseline}}}                     & \multicolumn{4}{c}{\small{\textbf{\ac{ourtool}}}}                    \\
                                 & \textbf{0\%}   & \textbf{25\%}  & \textbf{50\%}  & \textbf{75\%}  & \textbf{0\%}   & \textbf{25\%}  & \textbf{50\%}  & \textbf{75\%} \\
\midrule
			 SSDP  &  95.7 &  94.8 &  93.3 &  86.2 &  90.1 &  91.9 &  93.8 &  95.7 \\
            MSSQL  & 100.0 & 100.0 & 100.0 & 100.0 &  99.9 &  99.5 &  99.3 &  99.7 \\
          UDP-Lag  &  99.7 &  99.8 &  99.9 &  99.9 &  97.8 &  98.4 &  99.2 &  99.8 \\
             SNMP  & 100.0 & 100.0 & 100.0 & 100.0 & 100.0 &  99.8 &  99.8 &  99.9 \\
          Portmap  & 100.0 & 100.0 & 100.0 & 100.0 & 100.0 &  99.8 &  99.8 &  99.9 \\
          NetBIOS  & 100.0 & 100.0 & 100.0 & 100.0 & 100.0 &  99.8 &  99.8 &  99.9 \\
             LDAP  & 100.0 & 100.0 & 100.0 & 100.0 & 100.0 & 100.0 & 100.0 & 100.0 \\
              DNS  & 100.0 & 100.0 & 100.0 & 100.0 & 100.0 & 100.0 & 100.0 & 100.0 \\
\midrule
\small
\textbf{Average} &  \textbf{99.4} &  \textbf{99.3 }&  \textbf{99.1 }&  \textbf{98.3} &  \textbf{98.5} &  \textbf{98.6} & \textbf{99.0} &  \textbf{99.4} \\
\bottomrule
\end{tabular}
\end{adjustbox}
\end{table}

\subsubsection{Temporal evolution}
A comparison between the two approaches throughout the whole experiment is presented in Figures  \ref{fig:metrics-ssdp} and \ref{fig:metrics-udplag}. For a matter of space, we provide the plots of only two attacks, although the other attacks present similar behaviours. The plotted data consistently validates the average numbers presented earlier in this section, with no notable deviations observed over time. 

Besides the memory size and the packet rate, the performance of \ac{ourtool} is also influenced by the flow size and, as a consequence, by the flow rate. By collecting only $p=4$ packets/flow, the number of collected flows increases (and the \ac{sfnr} decreases) when the flow rate increases, as highlighted by the respective plots in the two Figures. On the contrary, the baseline approach is less affected by the flow rate, as it keeps on overwriting the old flows when the memory is full, resulting in approximately a stable \ac{sfnr} and number of collected flows.

\revised{
\subsection{Control Plane performance}
As described in Section \ref{sec:lucid}, the validation of \ac{ourtool} has been carried out using a modified configuration of \ac{lucid}, the \ac{ml}-based \ac{nids} operating in the control plane. This adaptation is essential to overcome the constraints posed by the \ac{p4} data plane. Notably, two key features are omitted (\textit{highest layer} and \textit{protocols}), and the packet count per flow $p$ is significantly reduced from $100$ to $4$. Under these settings, we measured an average FNR of approximately $0.013$ and an average accuracy and F1 Score of about $0.93$ on the CIC-DDoS2019 dataset across the experiments described in Section \ref{sec:emulated_scenario}.  

\begin{table*}[t!]
	\caption{\revised{Control plane performance}} 
	\label{tab:lucid_performance}
	\small 
	\centering 
	\begin{threeparttable}
		\begin{tabular}{lcccccccccc} \toprule[\heavyrulewidth]
			\textbf{Dataset}  & \multicolumn{2}{c}{\textbf{Accuracy}} & \multicolumn{2}{c}{\textbf{False Positive Rate}} & \multicolumn{2}{c}{\textbf{Precision}} & \multicolumn{2}{c}{\textbf{Recall}} & \multicolumn{2}{c}{\textbf{F1 Score}} \\ 
			 & \textbf{Offline} & \textbf{Online} & \textbf{Offline} & \textbf{Online} & \textbf{Offline} & \textbf{Online} & \textbf{Offline} & \textbf{Online} & \textbf{Offline} & \textbf{Online}\\ \midrule[\heavyrulewidth]
			ISCX2012 (IRC) & 0.9888 & 0.9612 & 0.0179 &0.0612 & 0.9827 & 0.9313 & 0.9952 & 0.9874 & 0.9889 & 0.9584\\ \midrule
			CIC2017 (LOIC) & 0.9967 & 0.9455 & 0.0059 & 0.0123 & 0.9939 & 0.9994 & 0.9994 & 0.9430 & 0.9966 & 0.9697 \\ \midrule
			CSECIC2018 (HOIC) & 0.9987 & 0.9922 & 0.0016 & 0.0154 & 0.9984 & 0.9978 & 0.9989 & 0.9978 & 0.9987 & 0.9956\\  \bottomrule[\heavyrulewidth]
		\end{tabular}
	\end{threeparttable}
\end{table*}

Nevertheless, we are keen to assess the impact of our adaptation on the \ac{lucid}'s overall performance. To achieve this objective, we have conducted further experiments in the emulated environment, replicating the three DDoS attacks used in the \ac{lucid}'s original paper.
The three attacks, generated using an IRC botnet and with well-known \ac{ddos} attacking tools such as LOIC \cite{loic} and HOIC \cite{hoic}, are part of three datasets provided by the University of New Brunswick, namely ISCX2012 \cite{ISCXIDS2012}, CIC2017 \cite{CICIDS2017} and CSECIC2018 \cite{CSECIC2018}.

The results presented in Table \ref{tab:lucid_performance} showcase the performance assessment of \ac{lucid} in segregating \ac{ddos} attacks from legitimate traffic under \textit{offline} and \textit{online} settings. The offline results are derived from \ac{lucid}'s original paper, where the evaluation was based on a static dataset of traffic flows represented by $11$ features, including \textit{highest layer} and \textit{protocols}, and with $p = 100$. Conversely, the online configuration aligns with the experimental setup detailed in earlier sections, featuring $9$ features and $p = 4$.

As reported in the table, the impact of reducing the feature matrix size from $100\times 11$ to $4\times 9$ is relatively insignificant in the case of the HOIC-based \ac{ddos} attack, with a noticeable but moderate increase in the FPR from $0.16\%$ to $1.5\%$. However, for the other attacks, we can observe a decrease of approximately $3\%$ in the F1 Score, indicating a general decline in overall performance, including a higher FPR in both instances.

It is worth mentioning that the \textit{online} results were obtained by training \ac{lucid} with just 9 features and 4 packets per flow. This highlights once again the remarkable adaptability of neural networks in adjusting their weights to the available input, underlining their flexibility and robustness in various configurations.

}

\section{Conclusion}\label{sec:conclusions}
In this paper, we have presented \ac{ourtool}, a solution based on P4 programmable data planes that enables the benefits of centralised intrusion detection while reducing the impact on control channel and hardware resources. \ac{ourtool} takes advantage of a probabilistic hashing data structure to carefully select the amount of information to be extracted from the network traffic and sent to the control plane, taking into consideration the traffic features required by the traffic classifier.

The key advantage of \ac{ourtool} over similar solutions resides in its ability to build a packet-level representation of traffic flows. This peculiarity allows  \ac{ourtool} to satisfy the requirements of state-of-the-art \ac{ml}-based \ac{nids}, which rely on a detailed representation of the traffic that goes beyond mere statistics. We have demonstrated that by using a counting Bloom filter to retain only the necessary information within the switch, \ac{ourtool} optimises the usage of stateful memory in the data plane, while reducing the chances of missed malicious flows due to lack of memory space.   

\bibliographystyle{IEEEtran}
\bibliography{main}

\end{document}